\documentclass[useAMS,usenatbib]{mn2e}
\usepackage{amsmath, amssymb}
\usepackage{graphicx}
\usepackage{wasysym}
\usepackage{subfigure}

\title[Dilution in elliptical galaxies]{Dilution in elliptical galaxies: Implications for the relation between metallicity, stellar mass and star formation rate}
\author[Yates \& Kauffmann]{Robert M. Yates$^{1}$\thanks{Email: robyates@mpa-garching.mpg.de} \& Guinevere Kauffmann$^{1}$ \\
$^{1}$ Max Planck Institut f$\ddot{u}$r Astrophysik, Karl-Schwarzschild-Str. 1, 85741, Garching, Germany}

\begin{document}
\date{Accepted ??. Received ??; in original form ??}
\maketitle\begin{abstract}
We investigate whether gradual dilution of the gas in some elliptical galaxies is the cause of a positive correlation between star formation rate (SFR) and gas-phase metallicity ($Z_{\textnormal{g}}$) at high stellar mass ($M_{*}$) in the local Universe. To do this, two classes of massive ($M_{*} \geq 10^{10.5} \textnormal{M}_{\textnormal{\astrosun}}$) galaxy are selected from the Sloan Digital Sky Survey (SDSS) and the Munich semi-analytic model of galaxy formation, \textsc{L-Galaxies}. The first class is selected by high specific star formation rates (sSFR) and high $Z_{\textnormal{g}}$, and the second class by low sSFR and low $Z_{\textnormal{g}}$. These criteria roughly distinguish disc-dominant galaxies from metal-poor, elliptical galaxies. In the semi-analytic model, the second class of galaxies obtain low sSFR and low $Z_{\textnormal{g}}$ due to \textit{gradual dilution} of the interstellar medium by accretion of metal-poor gas via infalling clumps and low-mass satellites. This occurs after a merger-induced starburst and the associated supernova feedback have quenched most of the original gas reservoir. A number of signatures of this evolution are present in these model galaxies at $z=0$, including low gas fractions, large central black holes, elliptical morphologies, old ages, and importantly, low $(Z_{\textnormal{g}}-Z_{*})$ indicating dilution after star formation. Remarkably, all of these properties are also found in low-sSFR, low-$Z_{\textnormal{g}}$, massive galaxies in the SDSS-DR7. This provides strong, indirect evidence that some elliptical galaxies are undergoing gradual dilution after a gas-rich merger in the local Universe. This dilution scenario also explains the \textit{positive} correlation between SFR and $Z_{\textnormal{g}}$ measured in high-$M_{*}$ galaxies, and therefore has consequences for the local fundamental metallicity relation (FMR), which assumes a weak \textit{anti}-correlation between SFR and $Z_{\textnormal{g}}$ above $\sim 10^{10.5} \textnormal{M}_{\textnormal{\astrosun}}$.
\end{abstract}


\section{Introduction} \label{sec:Introduction}
\LARGE{A }\normalsize considerable amount of attention in the recent literature has been devoted to studying the relation between stellar mass ($M_{*}$), star formation rate (SFR) and gas-phase metallicity ($Z_{\textnormal{g}}$) in galaxies. The $M_{*}$-SFR-$Z_{\textnormal{g}}$ relation is believed to be a stronger diagnostic of galactic chemical evolution than the simpler $M_{*}$-$Z_{\textnormal{g}}$ relation, as it provides constraints on the recent star formation, as well as the integrated star formation (i.e. $M_{*}$) and current $Z_{\textnormal{g}}$. However, despite this, there remain a number of possible explanations for the trends seen in this relation in the local Universe.

\citet{E08} found an anti-correlation between $Z_{\textnormal{g}}$ and both specific star formation rate (sSFR) and half-light radius at low $M_{*}$. This dependence was attributed to lower present-day star formation efficiencies in more compact galaxies, as rapid star formation at early times is believed to consume most of the cold gas in these systems. A flat fundamental plane relating $M_{*}$, SFR and $Z_{\textnormal{g}}$ was later found by \citet{LL10}, which extends unchanged out to $z\sim 3.5$. At the same time, a three-dimensional fundamental metallicity relation (FMR) was found by \citet{M10}. The FMR corrects for the observed anti-correlation between SFR and $Z_{\textnormal{g}}$ at low $M_{*}$ to provide a prediction of the metallicity of local galaxies with an expected 1$\sigma$ scatter of only $\sim 0.05$ dex. The SFR-$Z_{\textnormal{g}}$ dependence at low mass was assumed to be due to highly star-forming galaxies driving stronger galactic winds, which can efficiently remove metals from their small gravitational potential wells.

A study of SDSS-DR7 galaxies by \citeauthor{YKG12} (2012, hereafter YKG12) also found an anti-correlation between SFR and $Z_{\textnormal{g}}$ at low mass, but additionally a \textit{positive} correlation between these two properties at high mass. The key difference between the \citet{M10} and YKG12 studies was the metallicity diagnostic used -- the former took the average metallicity obtained from the $R_{23}$ (i.e. [O\textsc{ii}]+[O\textsc{iii}]/H$\beta$) and [N\textsc{ii}]/H$\alpha$ ratios, whereas the latter took the Bayesian metallicities provided by the MPA-JHU catalogue\footnote{available at; \textit{http://www.mpa-garching.mpg.de/SDSS/DR7}}, which are based on fitting six strong emission line fluxes to synthetic spectra (see \citealt{T04}). YKG12 argued that their choice of metallicity diagnostic is likely to be more robust for local, high-$Z_{\textnormal{g}}$ galaxies. This is because, a) the [N\textsc{ii}]/H$\alpha$ diagnostic is prone to \textit{under}-estimating the metallicity in this regime, due to saturation as the electron temperature drops below that required to easily excite the [N\textsc{ii}]$\lambda$6584 line. And b) the $R_{23}$ diagnostic, as calibrated by \citet{M08}, seems to \textit{over}-estimate the metallicity in this regime compared to the Bayesian technique by up to $\sim0.2$ dex, especially for lower-SFR galaxies.

Some recent observational studies support the findings of YKG12. For example, Lara-L\'{o}pez et al. (2012, in prep.) found a positive correlation between SFR and $Z_{\textnormal{g}}$ at high mass when using either a [N\textsc{ii}]/[O\textsc{ii}] or ([O\textsc{iii}]/H$\beta$)/([N\textsc{ii}]/H$\alpha$) diagnostic, demonstrating that it is not just Bayesian metallicities that produce such a trend. Similarly, \citet{AM12} have found a slight positive correlation at high mass when using a $R_{23}$ or [N\textsc{ii}]/H$\alpha$ diagnostic (separately). However, such a correlation is less clear when using the $T_{e}$ method, as their sample has only very few galaxies above $\textnormal{log}(M_{*}) = 10.5 \textnormal{M}_{\textnormal{\astrosun}}$ (see their figure 11). Additionally, metallicities derived using the $T_{e}$ method can be unreliable above $Z_{\textnormal{g}}\gtrsim 8.6$ due to local and global temperature gradients across galaxies (\citealt{S78a,St05}, but see \citealt{C13}). \citet{Z13} have found a correlation between SFR and dust extinction very similar to that found between SFR and $Z_{\textnormal{g}}$ by YKG12. As dust and metals are known to be produced and distributed in similar ways throughout galaxies (e.g. \citealt{D98}), these two correlations could share a common cause. Most recently, Kurk et al. (in prep.) have found a strong positive correlation between SFR and $Z_{\textnormal{g}}$ at high mass in a sample of LUCI/SINS galaxies at $z\sim 2$.

A possible explanation for this positive correlation was provided by YKG12, using the Munich semi-analytic model of galaxy formation, \textsc{L-Galaxies}. In the model, such a correlation is the consequence of gradual dilution of the interstellar medium (ISM) in low-SFR, massive galaxies by the accretion of metal-poor gas over several gigayears. Secular star formation is shut-down in these systems after a gas-rich merger, which produces a starburst, growth of the central black hole (BH), and ejection of gas via supernova feedback. Thereafter, the remaining gas is of too low density to continue forming stars, and the presence of `radio mode' AGN feedback suppresses cooling of hot gas from the circumgalactic medium (CGM). However, the accretion of metal-poor, cold gas clumps and low-mass satellites can still proceed (see \S \ref{sec:Dilution in elliptical galaxies}).

There should be a number of signatures at $z=0$ of this specific evolution. For example, YKG12 noted that they have larger-than-average central black holes for their mass. In this work, we identify a range of properties at $z=0$ in the model which are indicative of post-merger gradual dilution. We then utilise a wide array of publicly-available observational data to see if these signatures are also present in real low-SFR, low-$Z_{\textnormal{g}}$ galaxies at low redshift. If so, this would provide strong, indirect evidence that gradual dilution is also taking place in the real Universe.

In \S \ref{sec:The model sample}, we describe our model sample. In \S \ref{sec:ModelResults}, we present our model results, including a description of how dilution occurs in some massive, model galaxies. In \S \ref{sec:The observational sample}, we describe our observational sample and the methods used to obtain various galactic properties. In \S \ref{sec:ObsResults}, we present our observational results, and compare them to those from our model. In \S \ref{sec:Comparisons to other works} we discuss our results in the context of other studies. Finally, in \S \ref{sec:Conclusions}, we provide our conclusions.

\begin{figure*}
\centering
\includegraphics[totalheight=0.2\textheight, width=0.9\textwidth]{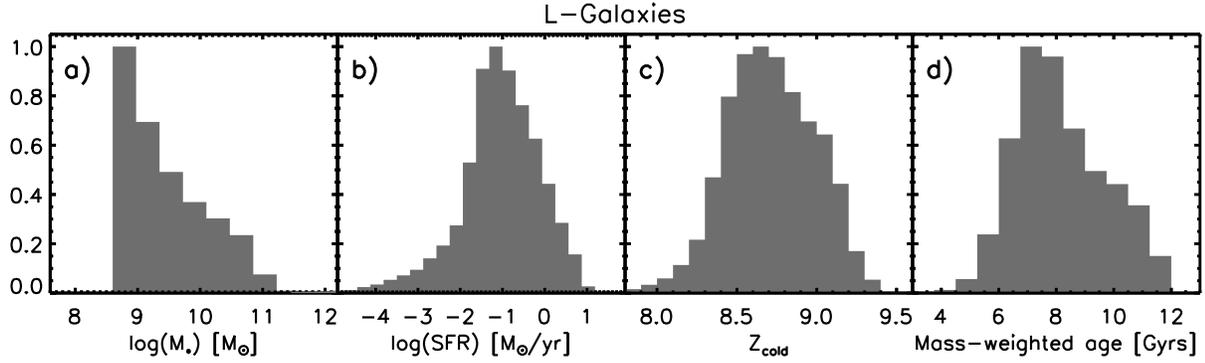}
\caption{The distribution in stellar mass (panel A), star formation rate (panel B), cold gas metallicity (panel C), and mass-weighted age (panel D) for our full model sample.}
\label{fig:LGals_SampleDists}
\end{figure*}

\section{The model sample} \label{sec:The model sample}
We form a sample of star-forming galaxies at $z=0$ from the Munich semi-analytic model of galaxy formation, \textsc{L-Galaxies} \citep{S01,DL04,S05,C06,DLB07,G10,G13,H13}. In the semi-analytic model, galaxy evolution is governed by the transfer of mass among the various galaxy components (central black hole, stellar bulge, stellar disc, ISM, CGM, halo stars, and ejecta reservoir), according to certain physical laws motivated by observations and simulations. In this work, we use outputs from the latest publicly-available version of \textsc{L-Galaxies} \citep{G10}, run on dark matter (DM) subhalo trees built from the \textsc{Millennium-II} N-body simulation \citep{BK09}. Our model sample was extracted from the Millennium Database\footnote{available at; \textit{http://www.g-vo.org/Millennium}} \citep{L06} provided by the German Astrophysical Virtual Observatory (GAVO). Galaxies were selected at $z=0$ only by stellar mass ($8.6\leq\textnormal{log(}M_{*}\textnormal{/M}_{\textnormal{\astrosun}})\leq11.5$), providing 64,523 model galaxies at $z=0$.\footnote{Type 2 galaxies (also known as `orphans'), whose DM sub-haloes have been stripped to below the DM subhalo resolution limit of $1.89\times 10^{8} \textnormal{M}_{\textnormal{\astrosun}}$, are not included in our analysis.}

\begin{figure}
\centering
\includegraphics[totalheight=0.26\textheight, width=0.38\textwidth]{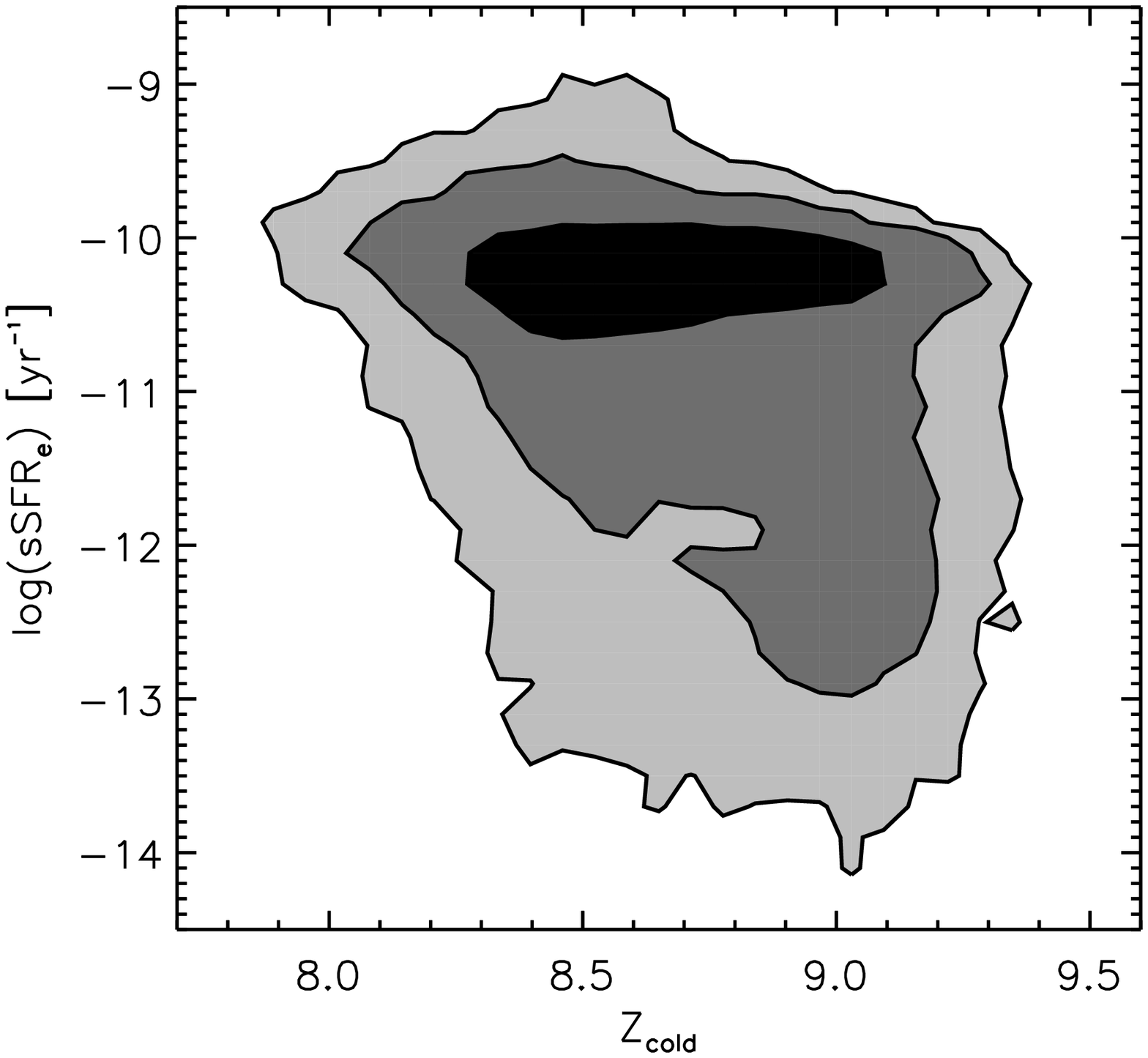} \\
\includegraphics[totalheight=0.26\textheight, width=0.38\textwidth]{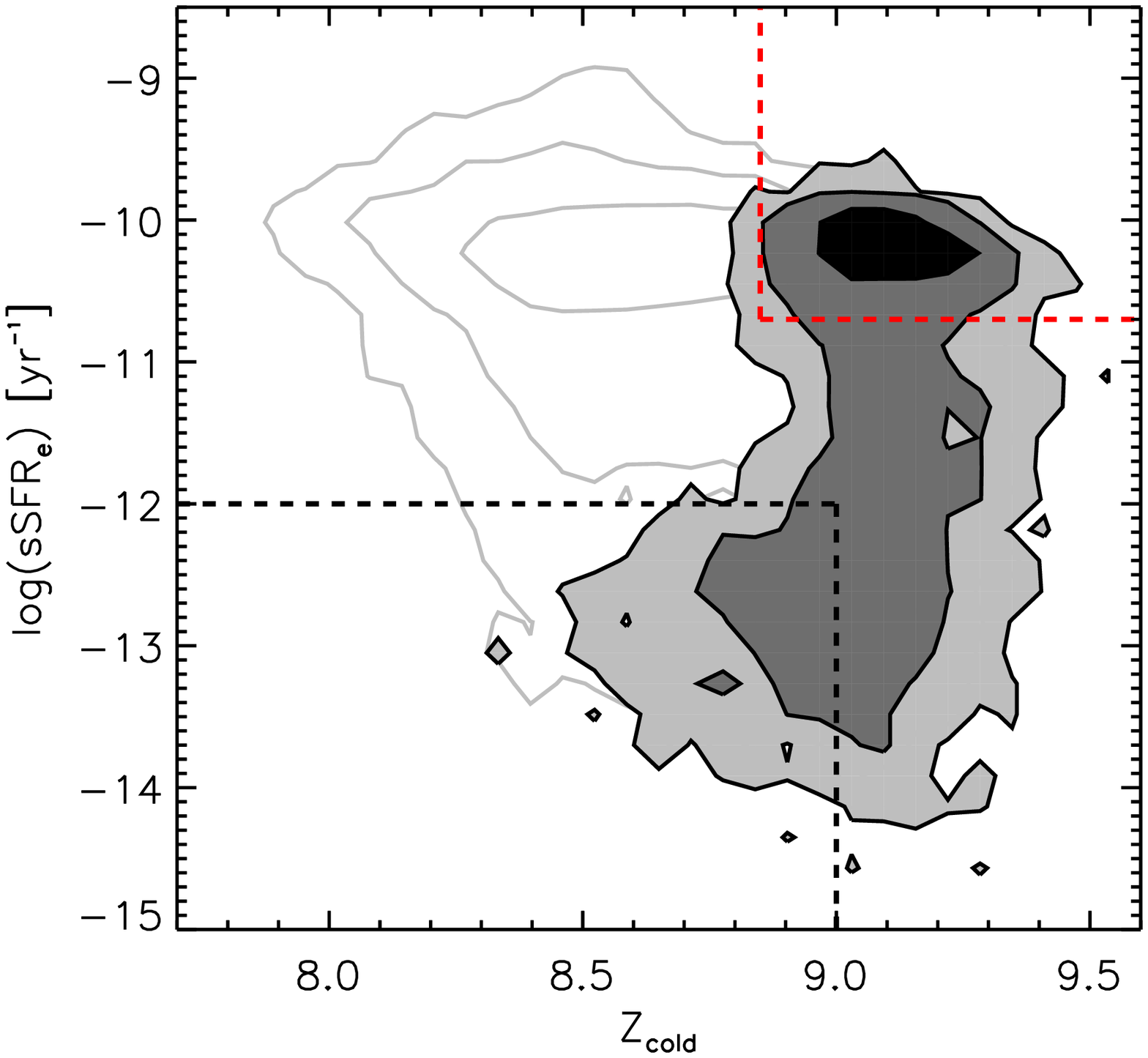}
\caption{\textit{Top panel}: The number density distribution for the whole model sample, in the sSFR-$Z_{\textnormal{cold}}$ plane. \textit{Bottom panel}: The number density distribution for galaxies with $\textnormal{log}(M_{*}) \geq 10.5 M_{\textnormal{\astrosun}}$ from the model sample, in the sSFR-$Z_{\textnormal{cold}}$ plane. The regions enclosed by the dashed red lines and dashed black lines define our enriching and diluting sub-samples, respectively. The contours of the distribution for the whole model sample are shown again in grey for reference.}
\label{fig:LGals_NumDensity}
\end{figure}

Fig. \ref{fig:LGals_SampleDists} shows the normalised $M_{*}$, SFR, $Z_{\textnormal{cold}}$ and mass-weighted age distributions of our model sample. The truncation at $\textnormal{log}(M_{*}) = 8.6 \textnormal{M}_{\textnormal{\astrosun}}$ is done to remove the large number of poorly resolved, very-low-mass galaxies in the model that are not present in our observational sample (\S \ref{sec:The observational sample}). We note that the focus of this work is on galaxies with $\textnormal{log}(M_{*}) \geq 10.5 \textnormal{M}_{\textnormal{\astrosun}}$.

\section{Model Results} \label{sec:ModelResults}
The top panel of Fig. \ref{fig:LGals_NumDensity} shows the number density distribution for our model sample in the sSFR-$Z_{\textnormal{cold}}$ plane. \citet{LL13} have also used the sSFR-$Z_{\textnormal{cold}}$ plane to study the relation between $M_{*}$, sSFR and $M_{\textnormal{HI}}$ in local galaxies. In our case, this plane is useful because it clearly separates the two classes of massive galaxy in which we are most interested (see \S \ref{sec:Two classes of massive galaxy}).

\begin{figure*}
\centering
\includegraphics[totalheight=0.58\textheight, width=0.8\textwidth]{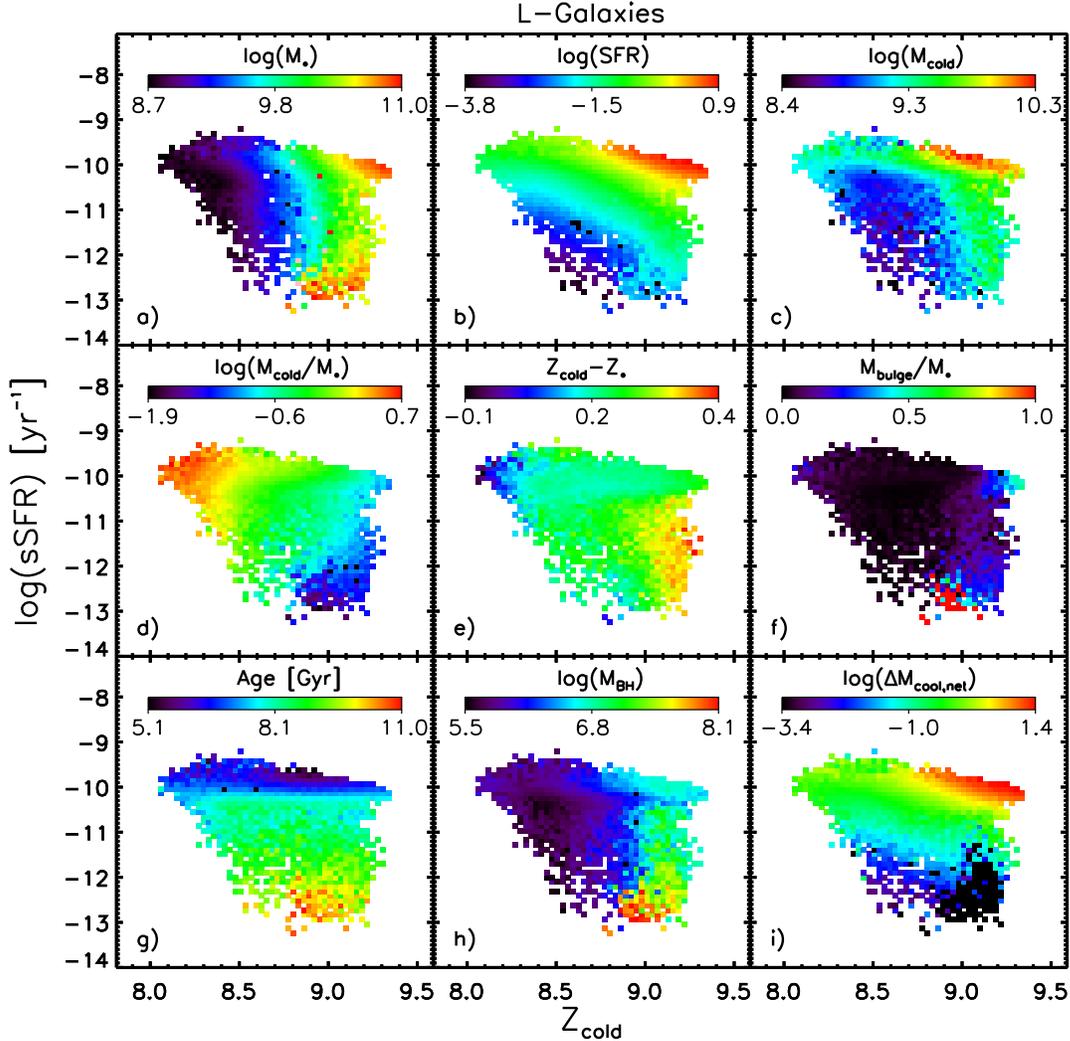}
\caption{Maps of the distribution of a number of properties in the sSFR-$Z_{\textnormal{cold}}$ plane for our full model sample. The property shown is stated at the top of each panel. The black (red) boxes in the top-left panel show the regions in which the diluting (enriching) sub-samples of massive ($M_{*} \geq 10^{10.5} \textnormal{M}_{\textnormal{\astrosun}}$) galaxies are located.}
\label{fig:LGals_SSFR-Z_maps}
\end{figure*}

Fig. \ref{fig:LGals_SSFR-Z_maps} shows `maps' of the full model sample in the same plane, with the colouring in each panel denoting a different physical property. Galaxies are binned by sSFR and $Z_{\textnormal{cold}}$, and only bins containing 10 or more galaxies are shown. There are clear trends in a number of properties for the sample as a whole. For example, $M_{*}$, $M_{\textnormal{cold}}/M_{*}$, and $M_{\textnormal{BH}}$ (panels A, C, and H) increase with gas-phase metallicity, whereas SFR, $M_{\textnormal{cold}}$, age, and net cooling rate (panels B, C, G and I) increase with sSFR. All of these trends are as we would expect from galaxy evolution in a hierarchical-merging, $\Lambda$CDM universe, where galaxies are typically expected to grow in mass and metallicity with time.

\subsection{Two classes of massive galaxy} \label{sec:Two classes of massive galaxy}
In order to study the SFR-$Z_{\textnormal{cold}}$ relation at high mass, we have selected two sub-samples of galaxies with $M_{*} \geq 10^{10.5}\textnormal{M}_{\textnormal{\astrosun}}$. The first contains 2,157 galaxies with $\textnormal{log sSFR} \geq -10.7\:\textnormal{yr}^{-1}$ and $Z_{\textnormal{cold}} \geq 8.85$. The second contains 711 galaxies with $\textnormal{log sSFR} \leq -12.0\: \textnormal{yr}^{-1}$ and $Z_{\textnormal{cold}} \leq 9.0$. For simplicity, we refer to these two sub-samples as \textit{enriching galaxies} and \textit{diluting galaxies}, respectively. The former population are typically undergoing an increase in $Z_{\textnormal{cold}}$ with time, whereas the latter population are typically undergoing a decrease in $Z_{\textnormal{cold}}$ with time (see YKG12). As we will see, they could equally be referred to as disc-dominated and bulge-dominated galaxies, young and old galaxies, or metal-rich and metal-poor galaxies. However, for the purposes of this work, we will label them by their typical net change in $Z_{\textnormal{cold}}$ at $z=0$.

The dashed red and black lines in the bottom panel of Fig. \ref{fig:LGals_NumDensity} define these two sub-samples. We can see that the enriching sub-sample (red) contains the high-$Z_{\textnormal{cold}}$ tip of the main galaxy distribution, whereas the diluting sub-sample (black) contains an extended distribution of galaxies that have fallen-off the main sequence at some point in the past.

\begin{figure}
\centering
\includegraphics[totalheight=0.25\textheight, width=0.38\textwidth]{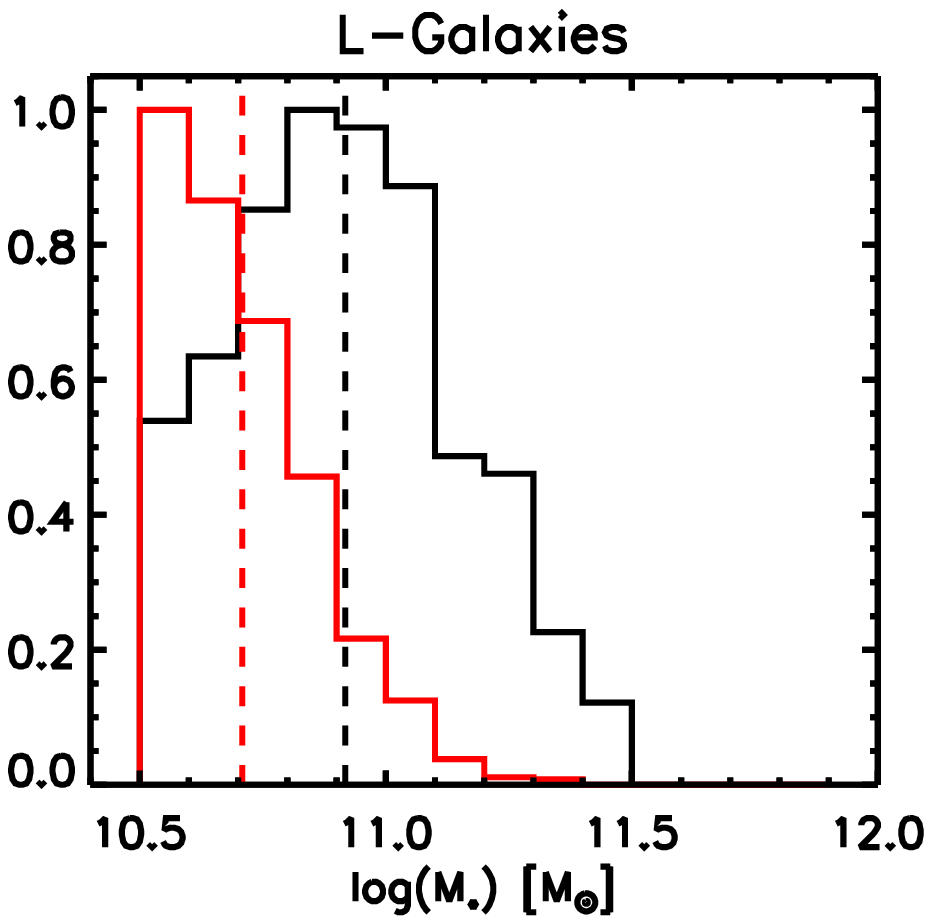}
\caption{The distribution of stellar mass for diluting galaxies (black) and enriching galaxies (red) from the model sample. Mean values are given by the dashed lines for each distribution.}
\label{fig:LGals_High-LowZ_Mdists}
\end{figure}

\begin{figure*}
\centering
\includegraphics[totalheight=0.35\textheight, width=0.58\textwidth]{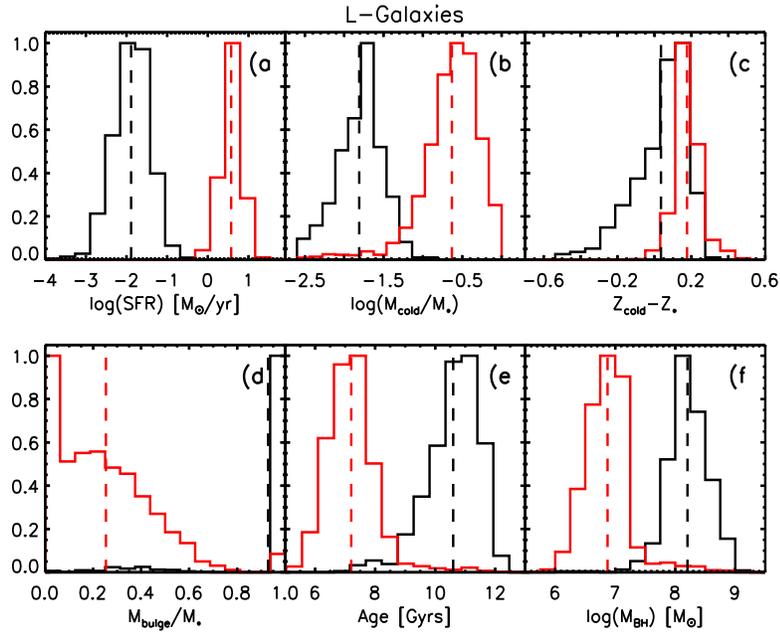}
\caption{The distribution of SFR (panel A), gas-to-stellar mass ratio (panel B), metallicity difference (panel C), bulge-to-total stellar mass ratio (panel D), mass-weighted age (panel E), and central black hole mass (panel F) for diluting galaxies (black) and enriching galaxies (red) from the model sample. Mean values are given by the dashed lines for each distribution.}
\label{fig:LGals_High-LowZ_hists}
\end{figure*}

\begin{figure}
\centering
\includegraphics[totalheight=0.25\textheight, width=0.38\textwidth]{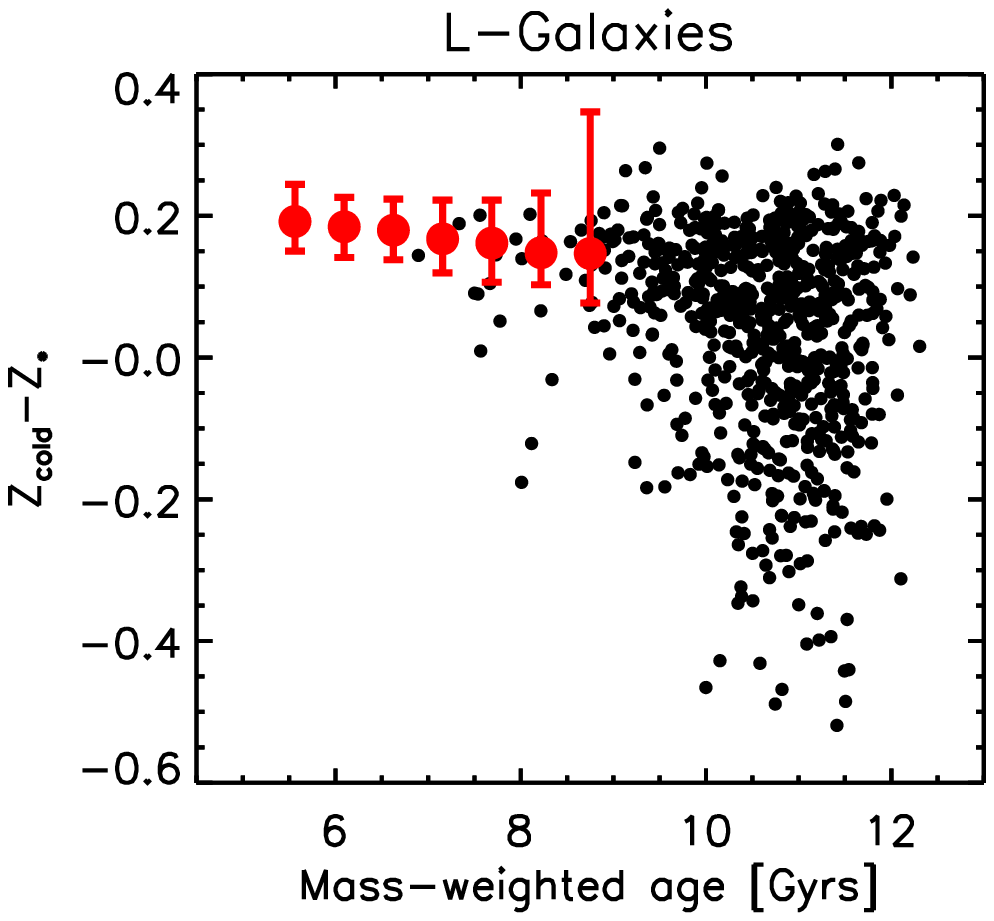}
\caption{The relation between mass-weighted age and $Z_{\textnormal{cold}}-Z_{*}$ for our model massive galaxies. Red points show the median $Z_{\textnormal{cold}}-Z_{*}$ in bins of age for high-SFR, high-$Z_{\textnormal{cold}}$, massive galaxies. Error bars indicate the 16th and 84th percentiles in each bin. Black points show individual low-SFR, low-$Z_{\textnormal{cold}}$, massive galaxies. Old low-SFR, low-$Z_{\textnormal{cold}}$, massive galaxies can have lower values of $Z_{\textnormal{cold}}-Z_{*}$ due to gradual dilution of their ISM in the absence of continuous star formation.}
\label{fig:LGals_age-Zdiff}
\end{figure}

We note here that the exact limits of the selection criteria are somewhat arbitrary. We have attempted to select massive galaxies with `typical' or enhanced star formation for our enriching sub-sample, and the low-SFR, low-$Z_{\textnormal{cold}}$ tail of the distribution for our diluting sub-sample. Small changes in the sizes of these regions do not affect any of our results. A cleaner sample of diluting galaxies could be selected by only choosing those systems with a negative change in $Z_{\textnormal{cold}}$ over the last few gigayears. 13.7 per cent of the `diluting' sample have undergone a slight net increase in $Z_{\textnormal{cold}}$ since $z=0.28$, and so can be considered contaminants, or at least early starters in an extended dilution process. Nonetheless, we only select galaxies by their $z=0$ properties, to provide a fairer comparison with our observational sample. A cleaner selection simply strengthens the dichotomy seen between the enriching and diluting sub-samples in the model.

Fig. \ref{fig:LGals_High-LowZ_Mdists} shows the stellar mass distribution for the model enriching (red) and diluting (black) galaxies. The mean $M_{*}$ is $\sim 0.2$ dex higher for the diluting sub-sample than the enriching sub-sample. This is because these galaxies tend to live in denser environments and have many more minor mergers (see \S \ref{sec:Dilution in elliptical galaxies}). However, we note that the local $M_{*}$-$Z_{\textnormal{g}}$ relation flattens-off above $\textnormal{log}(M_{*}) \sim 10.5\: \textnormal{M}_{\textnormal{\astrosun}}$ in both observations and our model, so this small difference in mean stellar mass does not imply enhanced $Z_{\textnormal{cold}}$ in diluting galaxies. In fact, these galaxies have been specifically selected to have low sSFR and low $Z_{\textnormal{cold}}$.

In Fig. \ref{fig:LGals_High-LowZ_hists}, we show histograms of the key physical properties of the enriching (red) and diluting (black) galaxies at $z=0$. We can clearly see that diluting galaxies have lower SFR (panel A), $M_{\textnormal{cold}}/M_{*}$ (panel B) and $Z_{\textnormal{cold}}-Z_{*}$ (panel C) than enriching galaxies, as well as higher $M_{\textnormal{bulge}}/M_{*}$ (panel D), older ages (panel E), and larger $M_{\textnormal{BH}}$ (panel F). All of these properties reflect the specific evolution that these galaxies have undergone -- a gradual dilution of the ISM after a merger-induced starburst that expelled gas via SN feedback and grew the central BH. Secular star formation and subsequent metal enrichment could not be resumed thereafter due to a) the small amount of remaining cold gas having a density below the threshold required for star formation, and b) the suppression of further cooling by AGN feedback.

In the case of $Z_{\textnormal{cold}}-Z_{*}$ (panel C), it is more precise to say that enriching galaxies form a tight distribution around $(Z_{\textnormal{cold}}-Z_{*}) \sim 0.17$, whereas diluting galaxies exhibit a strong tail down to low (even negative) values. This parameter is a useful diagnostic for dilution of the ISM \textit{after} the last bout of star formation, as low values of $Z_{\textnormal{cold}}-Z_{*}$ indicate a decrease in the gas-phase metallicity without a corresponding decrease in the stellar metallicity \citep{KE99}. Enriching galaxies are undergoing smooth, continuous star formation, and so have reached an equilibrium between their gas and stellar metallicities, whereas diluting galaxies are experiencing gradual dilution of the ISM at a greater rate than any star formation.

We should therefore expect that the value of $Z_{\textnormal{cold}}-Z_{*}$ in diluting galaxies anti-correlates with the time spent undergoing dilution. Fig. \ref{fig:LGals_age-Zdiff} demonstrates that this is indeed the case. Given that the mass-weighted age correlates with the dilution time (because no secular star formation occurs once dilution has started in these model galaxies), we can see that the oldest diluting galaxies (black points) have the lowest $Z_{\textnormal{cold}}-Z_{*}$.

\begin{figure*}
\centering
\includegraphics[totalheight=0.2\textheight, width=0.9\textwidth]{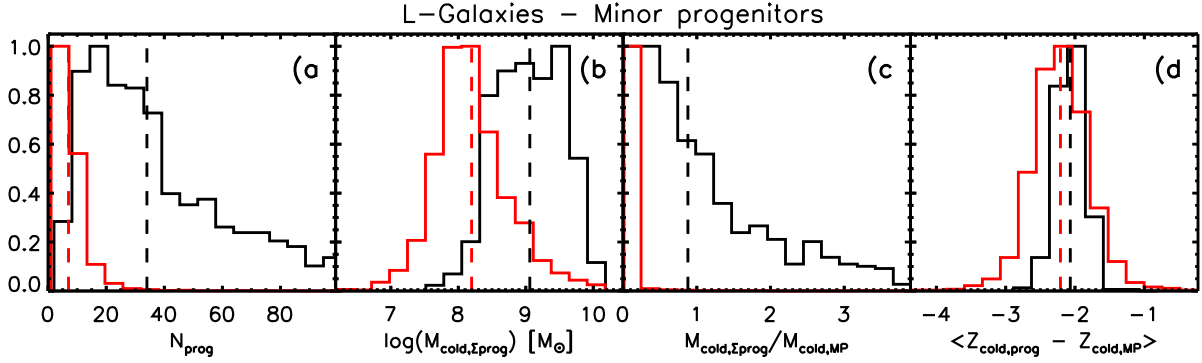}
\caption{Histograms illustrating the accretion of cold gas via minor mergers for enriching galaxies (red) and diluting galaxies (black) from $z=0.28$ to 0.0. \textit{Panel A:} The total number of minor progenitors (i.e. merging satellites). \textit{Panel B:} The total cold gas accreted via merging satellites. \textit{Panel C:} The ratio of total cold gas accreted to cold gas mass in the central galaxy at $z=0.28$. This parameter tells us how much a central galaxy's gas mass grew via cold accretion, with a value of 1.0 signifying that it doubled. \textit{Panel D:} The difference between the cold gas metallicity of merging satellites and that of the central galaxy. Dashed lines indicate the median values for panels A and C, and mean values for panels B and D.}
\label{fig:GasFracPaper_LGals_Prog_hists}
\end{figure*}

\subsection{Dilution in elliptical galaxies} \label{sec:Dilution in elliptical galaxies}
We will now discuss the \textit{type} of dilution that is occuring in our model diluting galaxies. The final panel in Fig. \ref{fig:LGals_SSFR-Z_maps} (panel I) shows the net cooling rate of gas, taking into account the suppression from AGN feedback. The \textit{gross} cooling rate of hot gas from the circumgalactic medium (CGM) is calculated following \citet{WF91} as,

\begin{align} \label{eqn:Gross_cooling_rate}
\dot{M}_{\textnormal{cool,gross}} = \bigg{\{} & \begin{array}{ll}
													(r_{\textnormal{cool}}/r_{\textnormal{vir}})\;(M_{\textnormal{hot}}/t_{\textnormal{dyn}}) & \textnormal{if } r_{\textnormal{cool}} \leqslant r_{\textnormal{vir}} \\
													M_{\textnormal{hot}}/t_{\textnormal{dyn}} & \textnormal{if } r_{\textnormal{cool}} > r_{\textnormal{vir}}
									 \end{array} \;\;,
\end{align}
where $r_{\textnormal{cool}}$ is the radius within which the cooling timescale is shorter than the dynamical time (which is given by $t_{\textnormal{dyn}}=r_{\textnormal{hot}}/V_{\textnormal{vir}}$), and $r_{\textnormal{hot}}$ is the radius out to which hot CGM gas extends in the system (this is the virial radius for central galaxies).

The AGN reheating rate is calculated following \citet{C06} as,

\begin{equation} \label{eqn:AGN_reheating_rate}
\dot{M}_{\textnormal{reheat,AGN}} = (0.2\:\dot{M}_{\textnormal{BH}}\:c^{2})/V_{\textnormal{vir}}^{2}\;\;,
\end{equation}
where the rate of accretion onto the central BH, $\dot{M}_{\textnormal{BH}}$, is given by,

\begin{equation} \label{eqn:BH_accretion_rate}
\dot{M}_{\textnormal{BH}} = \kappa\left(\frac{f_{\textnormal{hot}}}{0.1}\right)\left(\frac{M_{\textnormal{BH}}}{10^{8}/h\:\textnormal{M}_{\textnormal{\astrosun}}}\right)\left(\frac{V_{\textnormal{vir}}}{200\:\textnormal{km/s}}\right)^{3}\;\;,
\end{equation}
and the hot accretion efficiency is $\kappa=1.5\times10^{-3} \textnormal{M}_{\textnormal{\astrosun}}/\textnormal{yr}$.

The \textit{net} cooling rate is therefore,

\begin{equation} \label{eqn:Net_cooling_rate}
\dot{M}_{\textnormal{cool,net}} = \dot{M}_{\textnormal{cool,gross}} - \dot{M}_{\textnormal{reheat,AGN}}\;\;.
\end{equation}
For more details, see \citeauthor{G10} (2011, \S 3.9).

Panel I of Fig. \ref{fig:LGals_SSFR-Z_maps} shows us that such `diffuse' cooling is completely shut-down in diluting galaxies, due to strong AGN feedback, whereas it is still occurring in enriching galaxies.\footnote{Net cooling rates of zero have been set to the minimum positive cooling rate in the sample of $\textnormal{log}(\dot{M}_{\textnormal{cool,net}})=-3.4$ when plotting logarithmic values in this figure.} This indicates that the gradual dilution of diluting galaxies is \textit{not} due to diffuse cooling of CGM gas, contrary to the conclusion made by YKG12. Instead, we have found that this gradual dilution is due to the accretion of low-mass satellites ($\sim87$ per cent by mass) and cold gas clumps ($\sim13$ per cent by mass), carried-in by merging DM subhaloes. This merger-based accretion is \textit{not} affected by the radio jets emitting from the central BH, and so can occur despite the presence of AGN feedback.

Fig. \ref{fig:GasFracPaper_LGals_Prog_hists} illustrates the significance of this mode of accretion for diluting galaxies. Panel A shows that diluting galaxies tend to have a larger number of minor progenitors (i.e. mergers) since $z \sim 0.28$ than enriching galaxies. This means that, although the \textit{average} cold gas mass of a merging satellite is similar for both classes ($4.89\times10^{7} \textnormal{M}_{\textnormal{\astrosun}}$ for diluting galaxies, and $4.16\times10^{7} \textnormal{M}_{\textnormal{\astrosun}}$ for enriching galaxies), the \textit{total} mass of cold gas accreted is greater for diluting galaxies (panel B). Considering that diluting galaxies also have \textit{low} cold gas masses themselves (Fig. \ref{fig:LGals_SSFR-Z_maps}, panel B), such merger-based accretion can have a significant impact on their cold gas content by $z=0$. Panel C of Fig. \ref{fig:GasFracPaper_LGals_Prog_hists} illustrates this by showing the ratio of the total mass in cold gas accreted to the cold gas mass of the main progenitor at $z=0.28$ (i.e. the gross increase in $M_{\textnormal{cold}}$ due to mergers). For enriching galaxies, the median ratio is $\sim 1.0$ per cent, whereas for diluting galaxies, satellites add an additional 88 per cent in cold gas on average. This leads to enriching galaxies undergoing a net \textit{decrease} in $M_{\textnormal{cold}}$ of 11 per cent on average, whereas diluting galaxies undergo a net \textit{increase} in $M_{\textnormal{cold}}$ of 11 per cent on average, over the last 3.1 Gyr.

This accreted gas is more metal-poor than the cold gas in the central galaxy (panel D), and therefore causes the significant dilution of the gas phase over time in diluting galaxies. The fact that this gas comes in during many minor merger events means that the dilution is gradual rather than sudden, with $Z_{\textnormal{cold}}$ dropping by $\sim0.034$ dex per Gyr on average ($\sim0.048$ dex per Gyr on average for those galaxies that show a consistent decrease in $Z_{\textnormal{cold}}$ since $z=0.28$). This drop in $Z_{\textnormal{cold}}$ in low-SFR galaxies over time is the cause of the positive correlation between SFR and $Z_{\textnormal{cold}}$ at high mass in the model FMR at $z=0$.

\begin{figure}
\centering
\includegraphics[totalheight=0.29\textheight, width=0.38\textwidth]{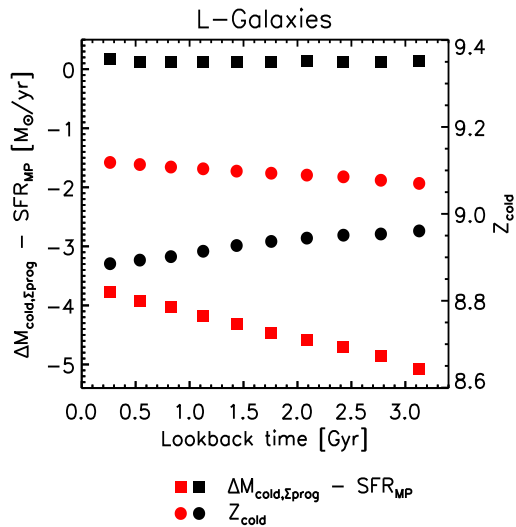}
\caption{The evolution of the \textit{median} accretion rate minus SFR (squares) and $Z_{\textnormal{cold}}$ (filled circles) with time, for enriching (red) and diluting (black) galaxies. The accretion rate, $\Delta M_{\textnormal{cold,}\Sigma\textnormal{prog}}$, is calculated as the mass of cold gas obtained through mergers per year. Enriching galaxies have negative $\Delta{}M_{\textnormal{cold,}\Sigma\textnormal{prog}}-\textnormal{SFR}_{\textnormal{MP}}$, due to low merger rates. Diluting galaxies have positive $\Delta{}M_{\textnormal{cold,}\Sigma\textnormal{prog}}-\textnormal{SFR}_{\textnormal{MP}}$, due to low SFRs.}
\label{fig:AccSFRDiff_Zcold_Evo}
\end{figure}

Another way to illustrate this evolution is to track the change in key galaxy parameters over time ($\textnormal{e.g.}$ YKG12, figs. 8 and 9). Fig. \ref{fig:AccSFRDiff_Zcold_Evo} shows the median $Z_{\textnormal{cold}}$ as a function of lookback time (filled circles), for enriching galaxies (red) and diluting galaxies (black). We can see $Z_{\textnormal{cold}}$ increases over cosmic time for enriching galaxies, and decreases over cosmic time for diluting galaxies. Fig. \ref{fig:AccSFRDiff_Zcold_Evo} also shows the evolution of the cold gas accretion rate minus SFR. This is a measure of the relative dilution/enrichment of the ISM; positive values indicate a net dilution (from infalling, metal-poor gas) and negative values indicate a net enrichment (from stars). We can see that enriching galaxies always have negative $\Delta{}M_{\textnormal{cold,}\Sigma\textnormal{prog}}-\textnormal{SFR}_{\textnormal{MP}}$. This is true even when including gas cooled from the CGM in the calculation. Diluting galaxies always have positive $\Delta{}M_{\textnormal{cold,}\Sigma\textnormal{prog}}-\textnormal{SFR}_{\textnormal{MP}}$. This shows again how these systems are gradually diluting their ISM over time. We note that the increase in $\Delta{}M_{\textnormal{cold,}\Sigma\textnormal{prog}}-\textnormal{SFR}_{\textnormal{MP}}$ for enriching galaxies is due to their average decline in SFR over time, reflecting the evolution of the cosmic SFR density \citep{M96}. The low SFRs and fairly constant merger rates of diluting galaxies ensures that their $\Delta{}M_{\textnormal{cold,}\Sigma\textnormal{prog}}-\textnormal{SFR}_{\textnormal{MP}}$ remains fairly constant over time.

It may be surprising that massive, bulge-dominant galaxies are undergoing minor mergers containing cold gas in the model, when the current understanding is that such systems grow in mass and size predominantly through dissipationless, minor mergers ($\textnormal{e.g.}$ \citealt{W76}; \citealt{NJO09}). However, both these pictures are consistent with each other, as the median SFR of these diluting galaxies is only $\sim0.06\:\textnormal{M}_{\textnormal{\astrosun}}/\textnormal{yr}$ since $z\sim0.28$, with only $0.7$ per cent of their present day stellar mass grown from forming new stars since then, on average. The accretion of stars via mergers over the same time is much more significant, contributing an average 13.7 per cent of the total stellar mass at $z=0$.

\begin{figure}
\centering
\includegraphics[totalheight=0.25\textheight, width=0.32\textwidth]{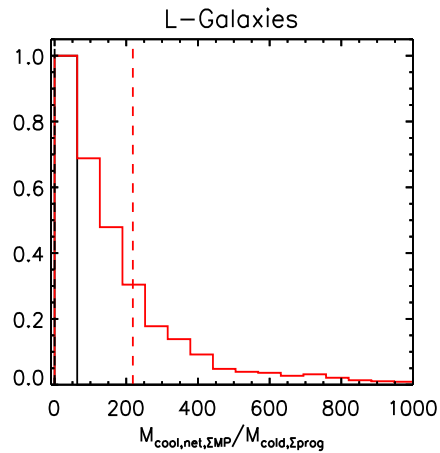}
\caption{The ratio of total gas mass cooled to total cold gas accreted for enriching galaxies (red) and diluting galaxies (black) from $z=0.28$ to 0.0. Dashed lines indicate the mean values. Cooling can be the dominant mechanism for obtaining cold gas in enriching galaxies, whereas accretion of satellites and gas clumps dominates for diluting galaxies.}
\label{fig:GasFracPaper_LGals_CoolvsAcc}
\end{figure}

Finally, in Fig. \ref{fig:GasFracPaper_LGals_CoolvsAcc} we show the ratio between `total gas mass cooled' and `total cold gas mass accreted' from $z\sim0.28$ to the present day. Cooling of CGM gas can clearly be a significant mode of obtaining cold gas in enriching galaxies, whereas it is negligible in diluting galaxies due to the presence of AGN feedback. This reflects the fact that accretion of cold gas via satellites and infalling gas clumps is the dominant mechanism for diluting galaxies in the model.

\textit{} \newline
In conclusion, we can say that there are a number of clear signatures of the dilution of some massive galaxies in the semi-analytic model that can be seen at $z=0$. These include: lower gas-to-stellar mass ratios, older ages, higher bulge-to-total stellar mass ratios, higher central BH masses, and lower $Z_{\textnormal{cold}}-Z_{*}$. We have also shown that metal-poor gas is accreted via minor merger events, rather than via diffuse cooling of hot gas from the CGM.

We now turn to the SDSS, to see if such features are also found in massive, low-sSFR, low-$Z_{\textnormal{g}}$ galaxies in the real Universe.

\begin{figure*}
\centering
\includegraphics[totalheight=0.2\textheight, width=0.9\textwidth]{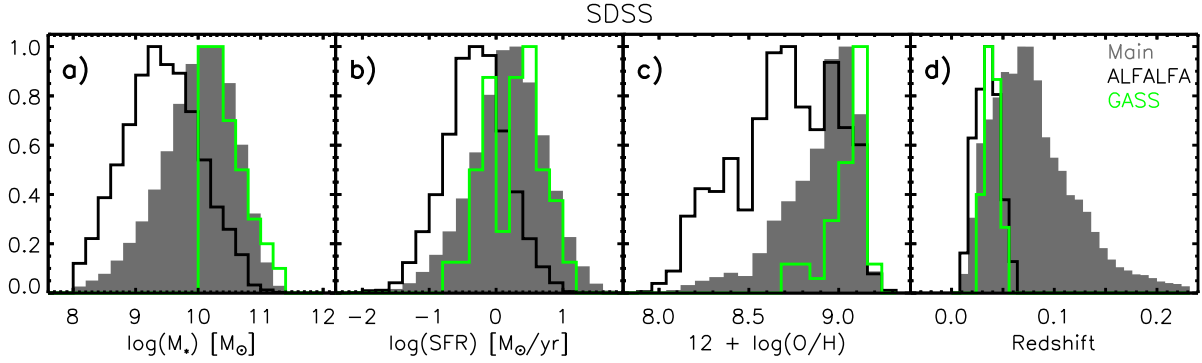}
\caption{The distribution in stellar mass (panel A), star formation rate (panel B), gas-phase metallicity (panel C), and redshift (panel D) for the Main observational sample (grey), only ALFALFA galaxies (black), and only GASS galaxies (green).}
\label{fig:SDSS_SampleDists}
\end{figure*}

\begin{figure}
\centering
\includegraphics[totalheight=0.26\textheight, width=0.38\textwidth]{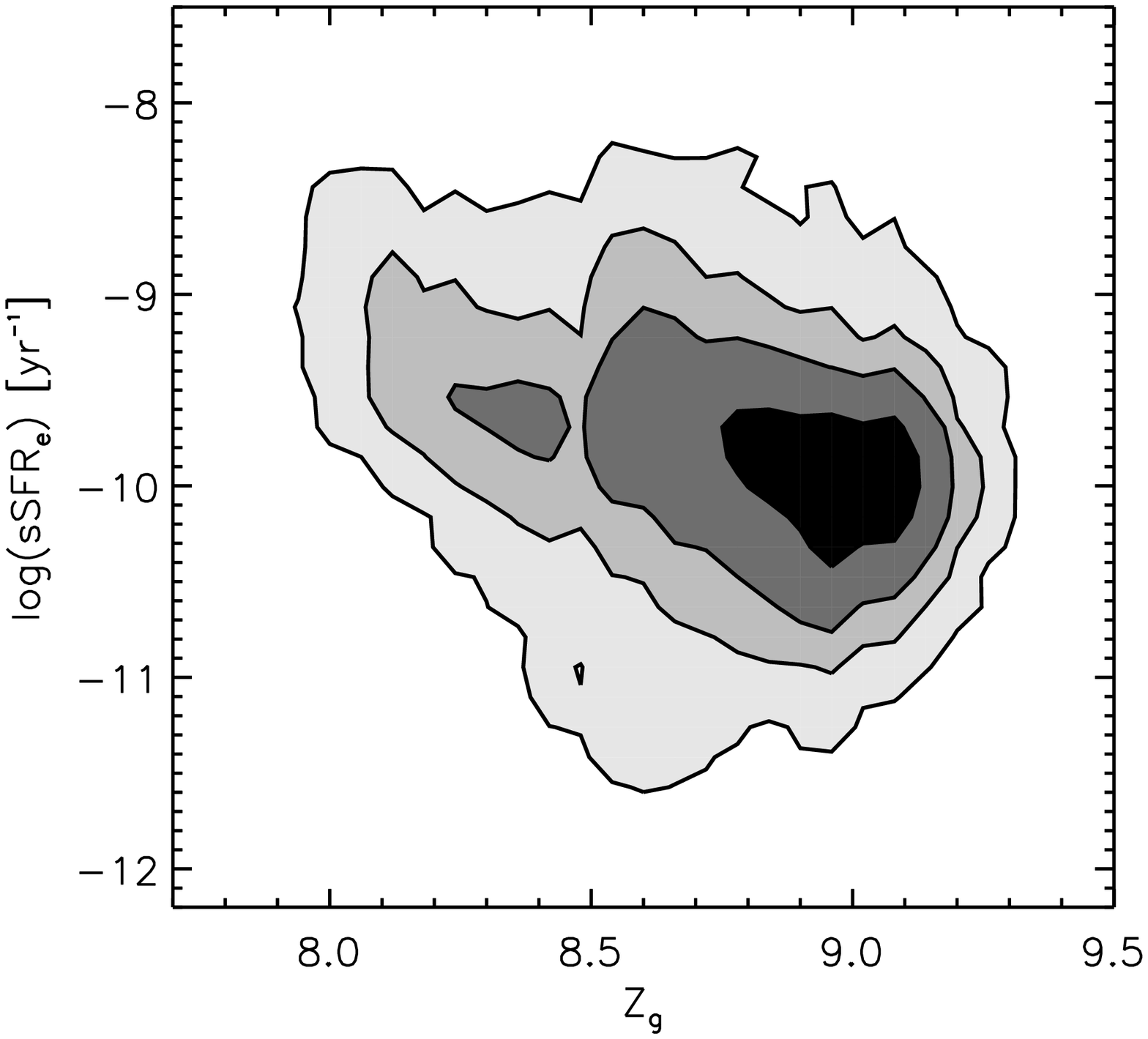} \\
\includegraphics[totalheight=0.26\textheight, width=0.38\textwidth]{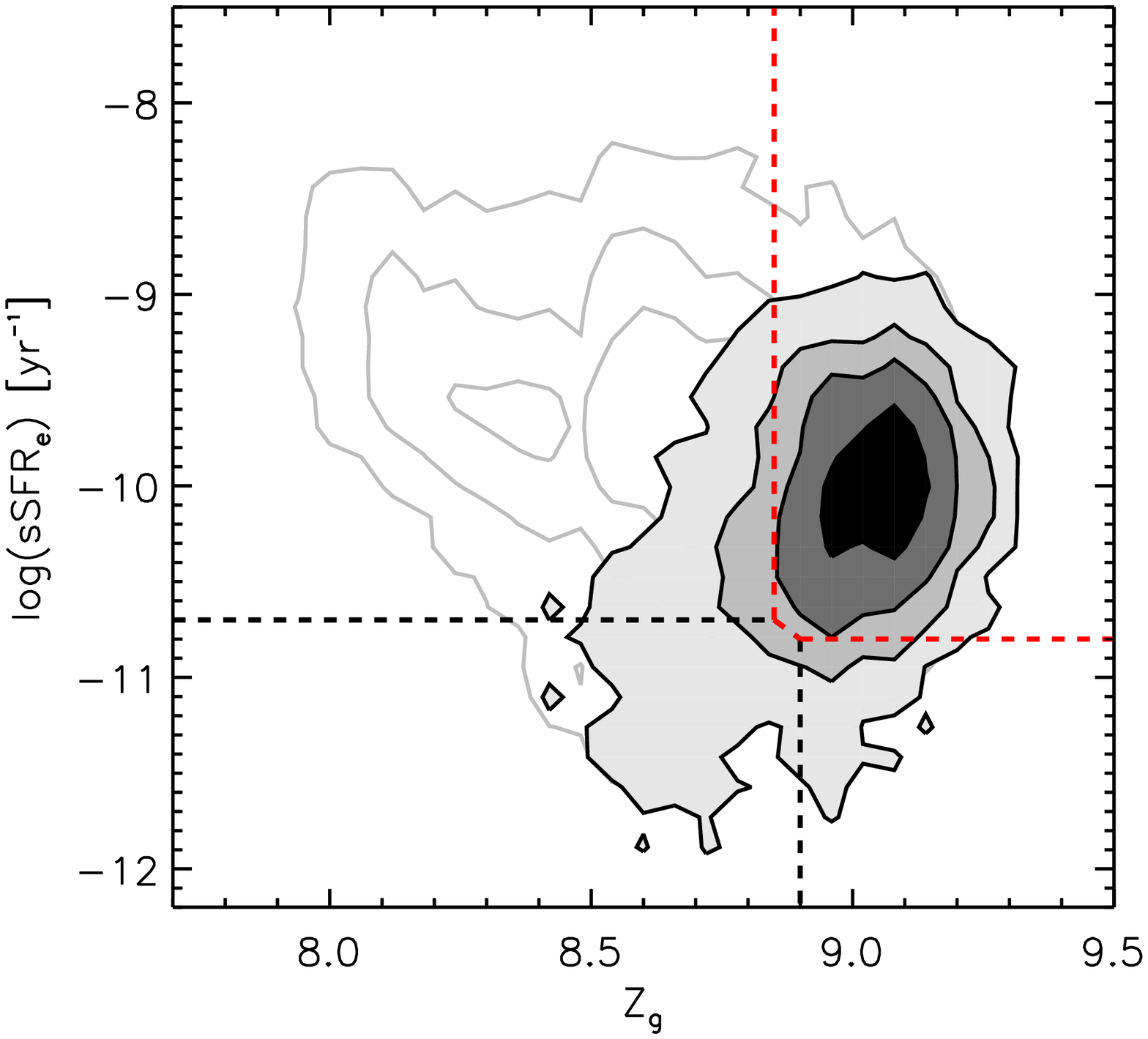}
\caption{\textit{Top panel}: The number density distribution for the main observational sample, in the sSFR-$Z_{\textnormal{g}}$ plane. {Bottom panel}: The number density distribution for galaxies with $\textnormal{log}(M_{*}) \geq 10.5 M_{\textnormal{\astrosun}}$ from the main observational sample, in the sSFR-$Z_{\textnormal{g}}$ plane. The regions enclosed by the dashed red lines and dashed black lines define our enriching and diluting sub-samples, respectively. The contours of the distribution for the whole main sample are shown again in grey for reference.}
\label{fig:SDSS_NumDensity}
\end{figure}

\section{The observational samples} \label{sec:The observational sample}
\subsection{Main sample} \label{sec:Main sample}
A main sample of 149,932 galaxies was selected from the SDSS-DR7\footnote{available at; \textit{http://www.mpa-garching.mpg.de/SDSS/DR7}}. 109,678 of these were obtained following the selection criteria of \citet{T04}, as outlined in section 2 of YKG12 for their Sample T2. We refer the reader to those works for further details. In brief, galaxies were selected to have r-band fibre-to-total light ratios $> 0.1$ and signal-to-noise ratios (SNR) of SNR(H$\alpha$, H$\beta$, [N\textsc{ii}]$\lambda 6584)>5$. AGN hosts were removed following \citet{K03c} using the \citet{BPT81} (BPT) diagram for galaxies with SNR([O$\textsc{iii}]\lambda5007)>3$. For galaxies with SNR([O$\textsc{iii}]\lambda5007)<3$, only those with $\textnormal{log}([\textnormal{N}\textsc{ii}]\lambda 6584/\textnormal{H}\alpha)<-0.4$ were retained, in order to remove low-ionization AGN hosts from the sample. Galaxies with a 1$\sigma$ spread $> 0.2$ in the likelihood distribution of the best-fitting value of $M_{*}$ and $Z_{\textnormal{g}}$ from \textsc{Cloudy} \citep{F98} were also removed. Finally, in order to be consistent with the original \citet{T04} sample, galaxies were also required to have $\sigma(m_{\textnormal{z}})<0.15$ sinh$^{-1}$(mag), $\sigma(\textnormal{H}\delta_{\textnormal{A}})<2.5$\AA, and $\sigma(\textnormal{D}_{n}4000)<0.1$.

An additional 40,254 galaxies were included for which $\sigma(M_{*})$ or $\sigma(Z_{\textnormal{g}}) > 0.2$, but that meet all the other requirements described above. The motivation for this is outlined in Appendix C of YKG12; these galaxies have a larger uncertainty in their stellar mass, due to errors in the SDSS u-band magnitudes which propagate through to the $M_{*}$ estimates. At $\textnormal{log}(M_{*})\geq 10.5$, these galaxies actually have estimates of $Z_{\textnormal{g}}$ well within the $\sigma(Z_{\textnormal{g}}) < 0.2$ requirement (see fig. C1 in YKG12), and 98 per cent also have $\sigma(M_{*}) < 0.3$. We therefore choose to include these galaxies in order to better probe the high-$M_{*}$, low-$Z_{\textnormal{g}}$ region of parameter space that we are interested in for this work. This brings our main sample to a total of 149,932 galaxies.

Stellar masses, total star formation rates and fibre-based, gas-phase metallicities are provided by the SDSS-DR7 catalogue. $M_{*}$ was obtained using fits to $u,g,r,i,z$ SDSS photometry, and have been corrected from a \citet{K01} to a \citet{C03} IMF. Total SFRs (also corrected to a Chabrier IMF) were corrected for dust using the \citet{C89} extinction law. SFR and $Z_{\textnormal{g}}$ were obtained by fitting galaxy emission-line spectra to a grid of synthetic spectra from \textsc{Cloudy} photoionisation models, as detailed by \citet{CL01}, and using the stellar population synthesis models of \citet{BC03}. For more information, see \citet{B04} and \citet{T04}. We note here that all our conclusions also hold when using a simpler, strong-line-ratio-based metallicity diagnostic (see \S \ref{sec:ObsResults}).

Fig. \ref{fig:SDSS_SampleDists} shows the normalised $M_{*}$, SFR, $Z_{\textnormal{g}}$ and $z$ distributions of our Main observational sample (grey). There are fewer low-$M_{*}$ and low-SFR galaxies than in our model sample (Fig. \ref{fig:LGals_SampleDists}). This is because such galaxies can be `lost' due to low luminosity or low SNR on the optical emission lines used for selection. This is not a significant issue in this work, as we focus on galaxies with $\textnormal{log}(M_{*}) \geq 10.5 \textnormal{M}_{\textnormal{\astrosun}}$.

The number density distribution of the whole Main sample is shown in the top panel of Fig. \ref{fig:SDSS_NumDensity}. In the bottom panel, the distribution for galaxies with $\textnormal{log}(M_{*})\geq 10.5$ is shown (see \S \ref{sec:Two classes of real massive galaxy}).

\subsection{HI-detected sample} \label{sec:HI-detected sample}
In order to assess the significance of gas fraction on the $M_{*}$-SFR-$Z$ relation, we formed a sub-sample containing those galaxies with direct detections of H\textsc{i} gas. There is an increasing amount of data available on the H\textsc{i} and H$_{2}$ contents of nearby massive galaxies, thanks to surveys such as ALFALFA \citep{Gi05}, GASS \citep{C10} and COLD GASS \citep{S11}. Also, scaling relations that provide an estimate of the gas fraction from other observable properties ($\textnormal{e.g.}$ \citealt{Z09,C12b,L12}) allow an analysis of the expected H\textsc{i} content for a much larger sample of galaxies (see \S \ref{sec:HI scaling relations}).

3,123 galaxies were found by cross-matching our main sample with the ALFALFA-$\alpha.40$ sample \citep{Ha11}.\footnote{available at; \textit{http://egg.astro.cornell.edu/alfalfa}} ALFALFA is a blind survey, detecting H\textsc{i} via the 21cm line within the footprint of the SDSS. In order to match to our main sample, we a) removed all ALFALFA objects with a heliocentric velocity ($v_{\textnormal{helio}}$) $< 3000 \textnormal{km/s}$. These are either high-velocity clouds within the Milky Way or galaxies for which redshifts cannot be accurately determined from $v_{\textnormal{helio}}$.\footnote{ALFALFA redshifts are determined by $v_{\textnormal{helio}}/c$, where $c$ is the speed of light in a vacuum.} b) removed all other ALFALFA objects which do not have the {\tt OCCode = I} flag. These are H\textsc{i} regions not associated with galaxies. c) cross-matched the right ascension (\textit{ra}), declination (\textit{dec}) and redshift ($z$) of our main sample with the remaining ALFALFA objects, allowing for maximum errors of $\sigma(ra, dec)=10$ arcsecs and $\sigma(z)=0.0003$.

The same maximum errors on \textit{ra}, \textit{dec} and $z$ were used to obtain 38 cross-matched galaxies from the GASS-DR1/DR2 samples \citep{C12b}.\footnote{available at; \textit{http://www.mpa-garching.mpg.de/GASS}} GASS is a targeted survey of $\sim 1,000$ known SDSS galaxies (232 of which have direct H\textsc{i} detections) of $M_{*} > 10^{10} \textnormal{M}_{\textnormal{\astrosun}}$, so no removal of intragalactic objects is required. Only galaxies with quality {\tt Q = 1} were retained. Right ascensions and declinations were obtained from the GASS data by decomposing the associated SDSS IDs (see Appendix A for details). Of these 38 galaxies, 9 are also found in our ALFALFA sub-sample. For these galaxies, we take the $M_{\textnormal{H\textsc{i}}}$ measurements obtained by GASS.

After cross-matching with these surveys, a total of 3,161 unique galaxies ($2.11$ per cent of our main sample) with direct $M_{\textnormal{H\textsc{i}}}$ measurements were obtained. The normalised $M_{*}$, SFR, $Z_{\textnormal{g}}$ and $z$ distributions for the ALFALFA sub-sample (black) and GASS sub-sample (green) are shown in Fig. \ref{fig:SDSS_SampleDists}, alongside the Main observational sample (grey).

\subsection{$Z_{*}$ sample} \label{sec:Zstar sample}
We also draw a sub-sample of galaxies for which stellar metallicities ($Z_{*}$) have been measured for the SDSS-DR4 \citep{G05}.\footnote{available at; \textit{http://www.mpa-garching.mpg.de/SDSS/DR4}} These galaxies were obtained using the same cross-matching requirements described in \S \ref{sec:HI-detected sample}. We use this sub-sample to obtain values of $Z_{\textnormal{g}}-Z_{*}$. As mentioned in \S \ref{sec:ModelResults}, low $Z_{\textnormal{g}}$ relative to $Z_{*}$ is indicative of dilution of the ISM by metal-poor infall after the last star-formation event.

We convert $Z_{*}$ from the SDSS-DR4 catalogue into units of $12+\textnormal{log(O/H)}$ as follows; $Z_{*,\textnormal{cat}}-\textnormal{log}(0.0134)+8.69$, where 0.0134 and 8.69 are the solar metallicity and oxygen abundance as determined by \citet{A09}, respectively. We note that alternative conversions using different solar values would only shift the amplitude of $Z_{\textnormal{g}}-Z_{*}$, and would not affect the relative values of this parameter for the two classes of massive galaxy considered in this work (see \S \ref{sec:Two classes of real massive galaxy}).

\citet{G05} point out that their stellar metallicity estimates are only reliably constrained for galaxies with a SNR per pixel of $\sim20.0$ or higher. Introducing such a cut reduces our $Z_{*}$ sample by 84 per cent (although it also \textit{strengthens} slightly the dichotomy in $Z_{\textnormal{g}}-Z_{*}$ for our two high-mass sub-samples, see \S \ref{sec:ObsResults}). Therefore, we instead choose a slightly weaker cut, selecting only those galaxies with a SNR per pixel of 14.8 (the mean value for the whole SDSS-DR4). Doing so reduces the $Z_{*}$ sample by only 59 per cent, to 24,275 galaxies, and produces very similar results to a sample using SNR per pixel $\geq 20.0$.

\subsection{NUV-r sample} \label{sec:NUV-r sample}
In order to obtain $M_{\textnormal{H\textsc{i}}}/M_{*}$ estimates via the H\textsc{i} scaling relation derived by \citet{C12b} (see \S \ref{sec:HI scaling relations}), we select 1,529 brightest cluster galaxies (BCGs) for which NUV-r colours have been measured by \citet{W10} (kindly provided by Jing Wang, priv. comm.). An additional 1,662 galaxies were obtained by cross-matching our Main sample with the Galaxy Evolution Explorer (GALEX) GR6 catalogue\footnote{available at; \textit{http://galex.stsci.edu/GR6/}}, matching objects by position and allowing for $\sigma$(\textit{ra},\textit{dec})$\leq 1$ arcsecond. A further 418 galaxies were obtained in the same way, by cross-matching our Main sample with the GALEX photometric data for objects in the Lockman Hole and Spitzer First Look Survey (FLS). This data was compiled for the Galaxy Multi-wavelength Atlas from Combined Surveys (GMACS) catalogue by \citet{J07a,J07b}.\footnote{available at; \textit{http://www.astro.columbia.edu/$\sim$bjohnson/\newline GMACS/catalogs.html}} The total number of galaxies in our NUV-r sample comes to 3,609.

\begin{figure}
\centering
\includegraphics[totalheight=0.26\textheight, width=0.38\textwidth]{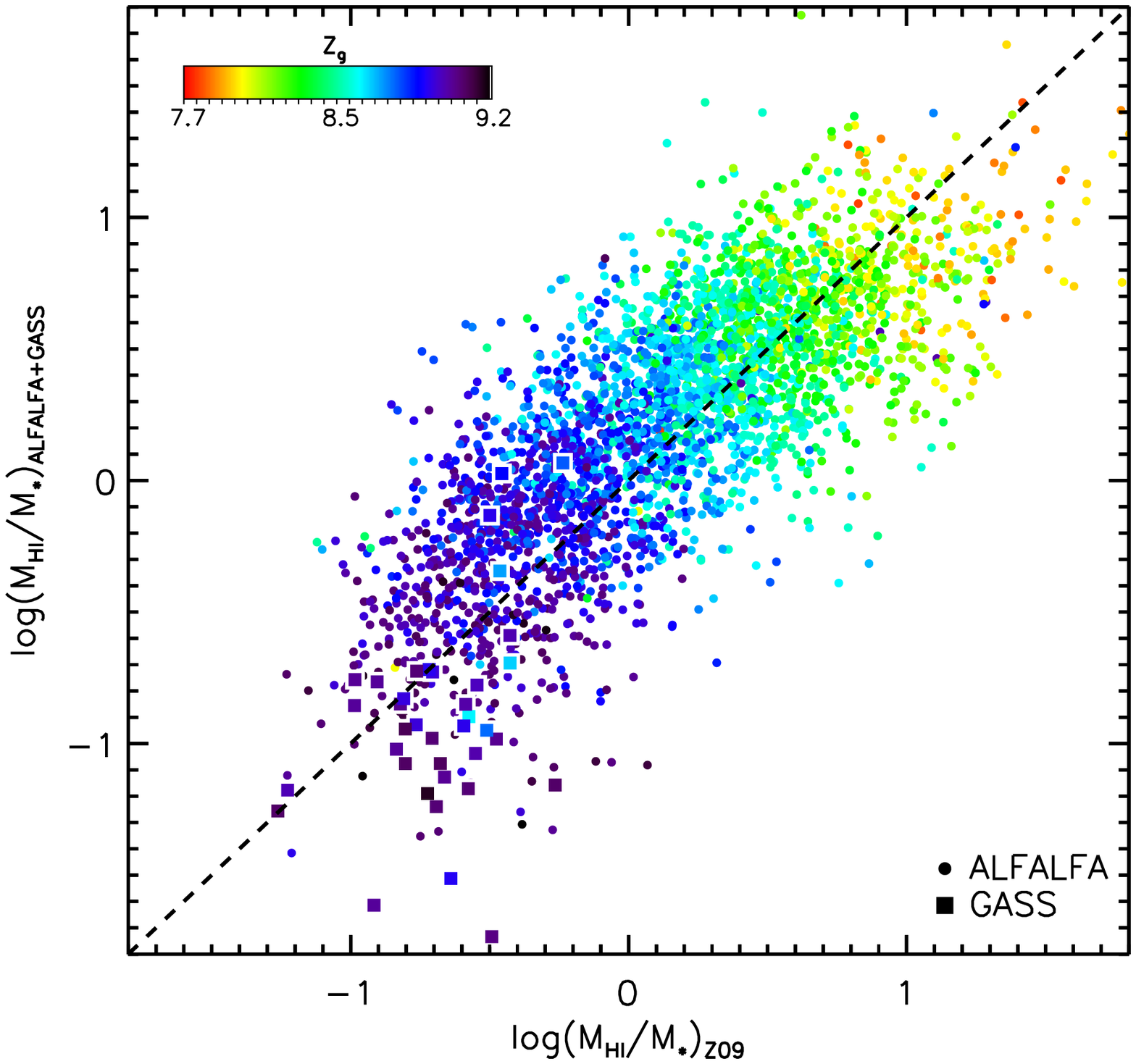} \\
\includegraphics[totalheight=0.26\textheight, width=0.38\textwidth]{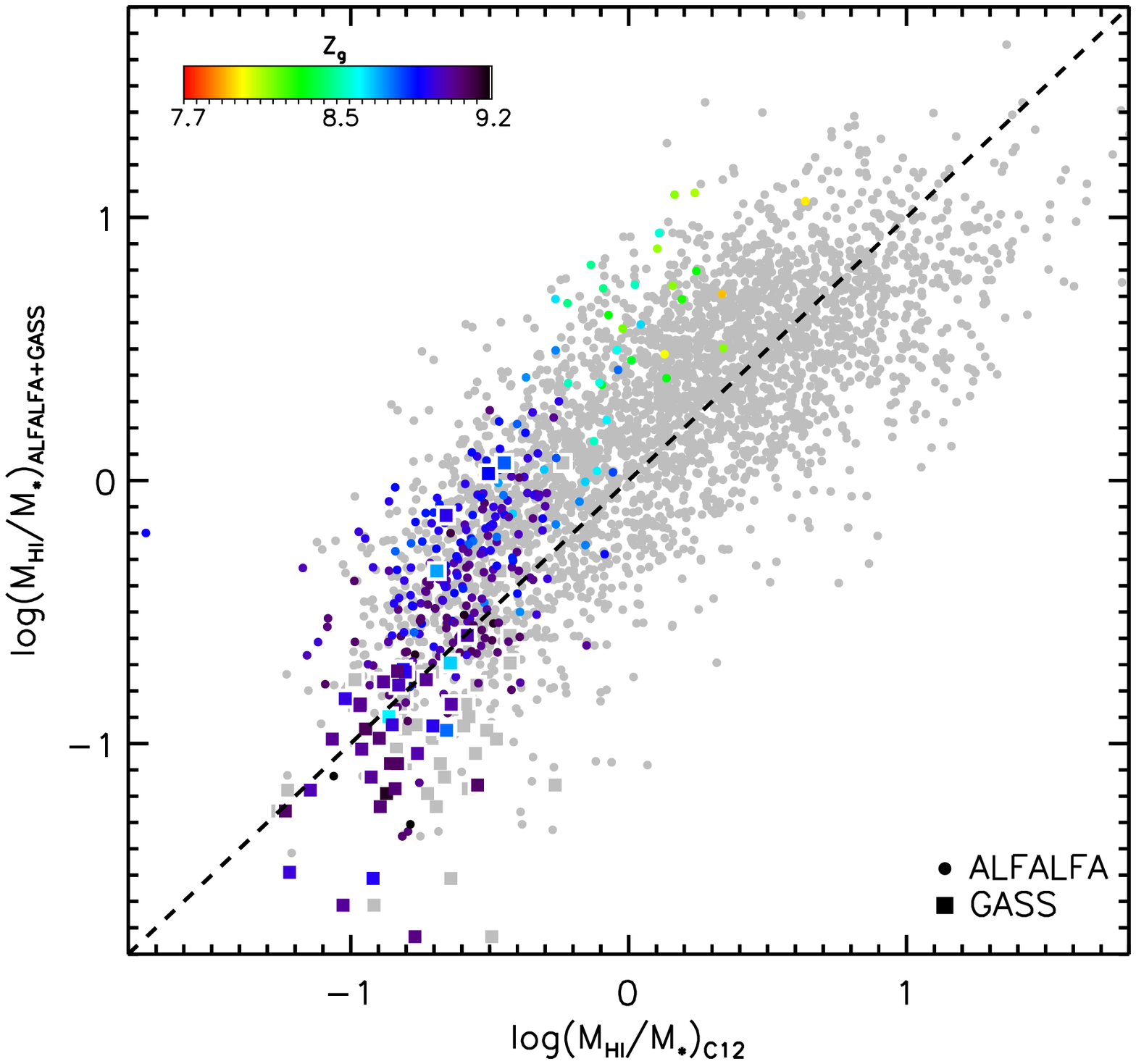} \\
\caption{\textit{Top panel:} A comparison between the H\textsc{i}-to-stellar-mass ratios obtained from the \citet{Z09} scaling relation and directly from ALFALFA (cirlces) and GASS (squares), for all 3,182 galaxies from our H\textsc{i}-detected sample. Data points are coloured by the gas-phase metallicity of each galaxy, as indicated by the colour bar. \textit{Bottom panel:} The same comparison, but between the \citet{C12b} scaling relation and ALFALFA (cirlces) and GASS (squares), for 228 galaxies from our NUV-r sample. For reference, the points from the top panel are also shown in grey.}
\label{fig:Rel-GF checks}
\end{figure}

\subsection{HI scaling relations} \label{sec:HI scaling relations}
As our H\textsc{i}-detected sample is only a small fraction ($2.1$ per cent) of our Main sample, we also utilise the H\textsc{i} scaling relation formulated by \citeauthor{Z09} (2009, hereafter Z09) to get $M_{\textnormal{H\textsc{i}}}$ estimates for all our galaxies. Z09 derived a mean relation between $M_{\textnormal{H\textsc{i}}}/M_{*}$, sSFR, and stellar surface brightness ($\mu_{*}$) for 800 SDSS-DR4 galaxies cross-matched with the HyperLeda H\textsc{i} catalogue \citep{P03}. Their relation is given by,

\begin{equation} \label{eqn:Z09}
\textnormal{log}(M_{\textnormal{H\textsc{i}}}/M_{*}) = -0.77\:\textnormal{log}(\mu_{*}) + 0.26\:\textnormal{log(sSFR)} + 8.53\;\;,
\end{equation}
where $\textnormal{log}(\mu_{*}) = \textnormal{log}(M_{*}/2\pi R_{50,i}^{2})$ and $R_{50,i}$ is the Petrosian i-band half-light radius (in arcsecs). All the properties required to estimate the gas-to-stellar-mass ratio from Eqn. \ref{eqn:Z09} are drawn from the SDSS-DR7 catalogue. Z09 also discuss the significance of gas fraction on the $M_{*}$-$Z_{\textnormal{g}}$ relation, and we compare our results to theirs in \S \ref{sec:The relation between M*, Zg and MHI}.

We first check that the Z09 scaling relation provides reasonable $M_{\textnormal{H\textsc{i}}}/M_{*}$ estimates for galaxies with direct H\textsc{i} detections from ALFALFA or GASS. This comparison is shown in the top panel of Fig. \ref{fig:Rel-GF checks}. We can see that, in general, the agreement is good, although the scatter is large. However, in detail, the Z09 scaling relation seems to predict larger $M_{\textnormal{H\textsc{i}}}/M_{*}$ than is measured by GASS (squares).

The GASS survey was specifically designed to observe galaxies until either an H\textsc{i} detection is made or a gas fraction limit of 0.015 is determined \citep{C12b}. This allows detections down to much lower H\textsc{i} masses than is possible by ALFALFA, which has an exposure time per galaxy of around a factor of ten smaller than GASS.

We check if the disparity at low $M_{\textnormal{H\textsc{i}}}/M_{*}$ is specific to the Z09 scaling relation by also comparing direct H\textsc{i} measurements to the $M_{\textnormal{H\textsc{i}}}/M_{*}$ estimates obtained from the \citet{C12b} scaling relation. This relation is calibrated using GASS galaxies and uses NUV-r colour rather than sSFR derived from optical emission lines. \citet{C10,C12b} found that massive, H\textsc{i}-detected galaxies form a flat, 2-dimentional plane in the $(M_{\textnormal{H\textsc{i}}}/M_{*})$-$\mu_{*}$-(NUV-r) parameter space, which can be well described by,

\begin{equation} \label{eqn:C12b}
\textnormal{log}(M_{\textnormal{H\textsc{i}}}/M_{*}) = -0.338\:\textnormal{log}(\mu_{*}) - 0.235\:\textnormal{(NUV-r)} + 2.908\;\;.
\end{equation}
This comparison is shown in the bottom panel of Fig. \ref{fig:Rel-GF checks} for 228 galaxies from our NUV-r sample that also have direct H\textsc{i} measurements. The \citet{C12b} relation seems to provide a similar range of $M_{\textnormal{H\textsc{i}}}/M_{*}$ estimates as the Z09 relation (grey points) for high-$Z_{\textnormal{g}}$ galaxies. This suggests that the larger scatter found at low $M_{\textnormal{H\textsc{i}}}/M_{*}$ is intrinsic to the difficulty in obtaining good 21 cm measurements for galaxies of such low gas fraction. We will show in \S \ref{sec:ObsResults} that both the direct $M_{\textnormal{H\textsc{i}}}/M_{*}$ estimates, and the two scaling relations described here, indicate larger gas fractions in enriching galaxies than in diluting galaxies.

We also note that the \citet{C12b} relation seems to under-estimate the gas-to-stellar-mass ratio for low-$Z_{\textnormal{g}}$, high-$M_{\textnormal{H\textsc{i}}}/M_{*}$ galaxies, compared to direct measurements (green and yellow points in the bottom panel of Fig. \ref{fig:Rel-GF checks}). \citet{L12} have shown that estimators which do not take account of colour gradients in galaxies can under-estimate $M_{\textnormal{H\textsc{i}}}/M_{*}$ in such gas-rich systems. They propose a new estimator, which includes the $g-i$ colour gradient ($\Delta_{g-i}$) to account for this effect. However, such a correction is not required in this work, as we choose to focus on galaxies with relatively high-$Z_{\textnormal{g}}$ and low-$M_{\textnormal{H\textsc{i}}}/M_{*}$.

\begin{figure*}
\centering
\includegraphics[totalheight=0.58\textheight, width=0.8\textwidth]{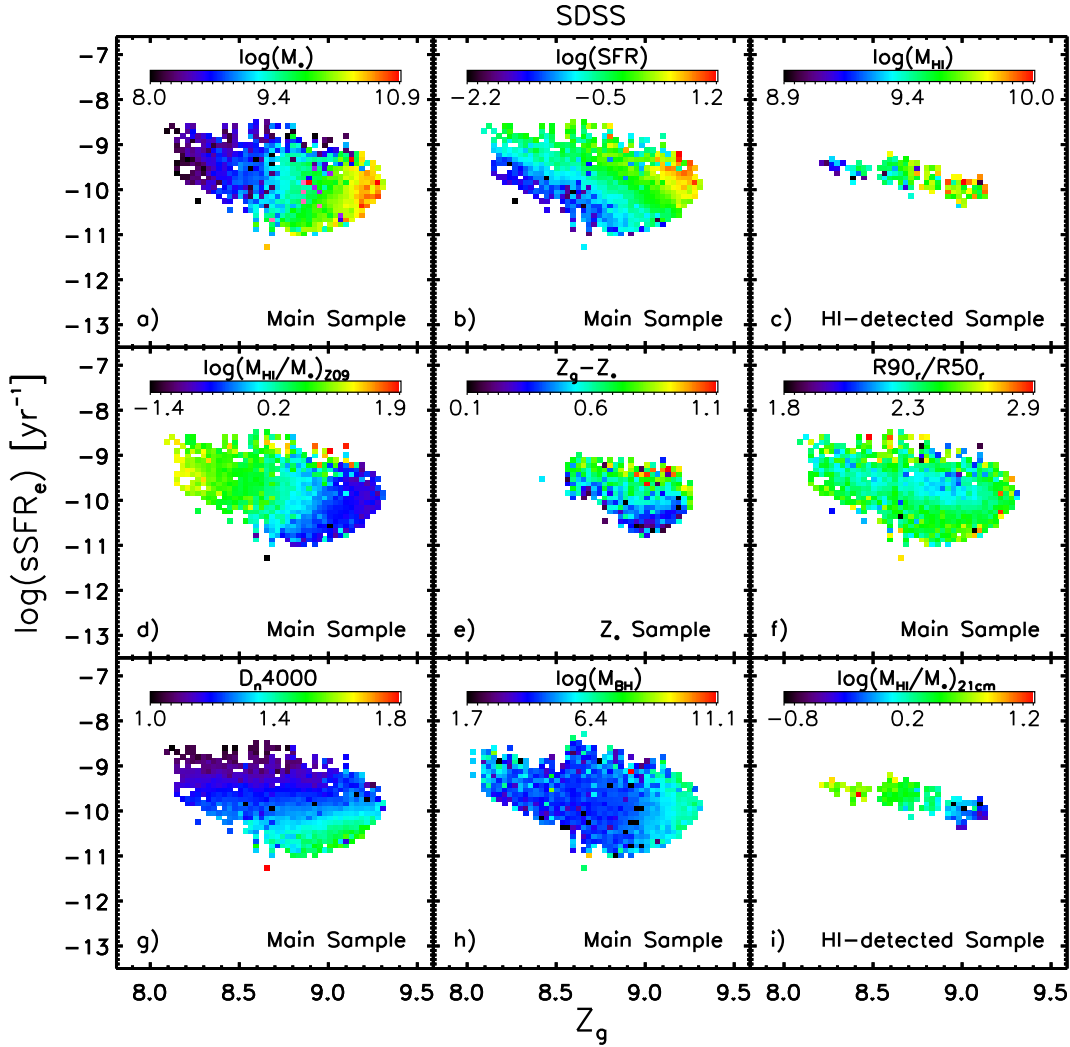}
\caption{Maps of the distribution of a number of properties in the sSFR-$Z_{\textnormal{g}}$ plane for the observational samples. The property shown is stated at the top of each panel. The black (red) boxes in the top-left panel show the regions in which the diluting (enriching), massive ($M_{*} \geq 10.5 \textnormal{M}_{\textnormal{\astrosun}}$ galaxies are located.}
\label{fig:SDSS_SSFR-Z_maps}
\end{figure*}

\section{Observational Results} \label{sec:ObsResults}
Fig. \ref{fig:SDSS_SSFR-Z_maps} shows `maps' of the observational samples in the sSFR-$Z_{\textnormal{g}}$ plane, in the same way as done for the model sample in Fig. \ref{fig:LGals_SSFR-Z_maps}. Galaxies are again binned by sSFR and $Z_{\textnormal{g}}$, and only bins containing 10 or more galaxies are plotted. For this figure, each galaxy is weighted by $1/V_{\textnormal{max}}$, the inverse of the maximum volume within which a galaxy of that r-band magnitude could be observed by the SDSS. This gives a greater weighting to faint, low-mass galaxies, to account for Malmquist bias.

The first thing we note when comparing Figs. \ref{fig:LGals_SSFR-Z_maps} and \ref{fig:SDSS_SSFR-Z_maps} is the different regions of parameter space covered. For example, the median sSFR for the model sample ($\textnormal{sSFR}=10^{-10.2}$) is lower than that of the observational sample ($\textnormal{sSFR}=10^{-9.82}$). This is because galaxies, particularly those with low $M_{*}$, tend to form stars too efficiently at high $z$ in the semi-analytic model. This means that lower SFRs are required at low $z$ in order to fit the $z=0$ stellar mass function ($\textnormal{e.g.}$ \citealt{G10}). \citet{H13} have since addressed this problem, by allowing material ejected from model galaxies to return to the ISM over longer periods of time, increasing their SFRs at low $z$ (see their fig. 9).

Also, there is a greater fraction of low-sSFR galaxies in the model sample than in the observational sample. This is likely due to the difficulty in obtaining SFR, metallicity and gas mass estimates for such galaxies in the real Universe, which will have intrinsically weaker emission line strengths, with lower SNR. Galaxies in this region of parameter space are also more likely to host AGN, as we believe them to be post-merger systems with large black holes (according to their model analogues), and so many may have been removed via the AGN cut.

Nonetheless, clear similarities can still be seen between the model and observational samples. Fig. \ref{fig:SDSS_SSFR-Z_maps} shows that low-mass galaxies have lower SFRs (panel B), lower H\textsc{i} masses (estimated via the 21cm line measurements of ALFALFA and GASS, panel C), higher gas-to-stellar mass ratios (using both the Z09 scaling relation, panel D, and direct estimates, panel I), lower metallicity differences (panel E), larger concentration indices (measured as the ratio of radius containing 90 per cent of the Petrosian r-band light to the half-light radius, panel F), older ages (inferred from the time since the last starburst via D$_{n}$4000, panel G), and lower-mass central black holes (inferred from velocity dispersions via the \citealt{Gr10} combined $M_{\textnormal{BH}}$-$\sigma$ relation, panel H). All of these trends are also found in our semi-analytic model, \textsc{L-Galaxies} (see Fig. \ref{fig:LGals_SSFR-Z_maps}).

We have also checked the stability of our results to changes in the selection criteria. When increasing the minimum SNR(H$\alpha$, H$\beta$, [N\textsc{ii}]) to 10, the low-sSFR edge of the galaxy population is `trimmed' slightly, increasing the median SFR of the whole Main sample by $\sim 0.02$ dex. Conversely, decreasing the maximum redshift to 0.1 removes some high-sSFR galaxies, decreasing the median SFR of the whole Main sample by $\sim 0.18$ dex. Finally, increasing the minimum fibre-to-total light ratio to 0.35 mainly removes low-redshift galaxies, as these tend to have larger apparent sizes, and reduces the Main sample to 32,550 objects. Despite these changes to the size and extremities of the galaxy population, the general trends described above are all unaffected by such changes to the selection criteria. The main conclusions for our high-$M_{*}$ sub-samples are also robust to these changes (see \S \ref{sec:Two classes of real massive galaxy}).

\begin{figure}
\centering
\includegraphics[totalheight=0.25\textheight, width=0.38\textwidth]{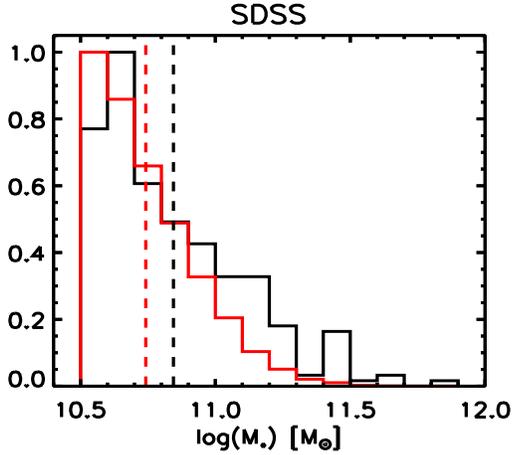}
\caption{The distribution of stellar mass for diluting galaxies (black) and enriching galaxies (red) from the Main sample. Mean values are given by the dashed lines for each distribution.}
\label{fig:SDSS_High-LowZ_Mdists}
\end{figure}

\begin{table*}
\centering
\begin{tabular}{l|cccccc|cccccc}
\hline \hline
 & \multicolumn{6}{c}{\textbf{Diluting galaxies}} & \multicolumn{6}{c}{\textbf{Enriching galaxies}} \\
\hline
 & \multicolumn{3}{c}{Model} & \multicolumn{3}{c}{Observations} & \multicolumn{3}{c}{Model} & \multicolumn{3}{c}{Observations} \\
 & \multicolumn{3}{c}{\textit{711 galaxies}} & \multicolumn{3}{c}{\textit{269 galaxies}} & \multicolumn{3}{c}{\textit{2157 galaxies}} & \multicolumn{3}{c}{\textit{28838 galaxies}} \\
\hline
  & Min & Median & Max & Min & Median & Max & Min & Median & Max & Min & Median & Max \\
\hline
log(sSFR) [yr$^{-1}$] & -15.4 & -12.7 & -12.0 & -12.8 & -11.0 & -10.7 & -10.7 & -10.1 & -9.1 & -10.8 & -9.9 & -8.4 \\
$Z_{\textnormal{g}}$ & 8.1 & 8.9 & 9.0 & 8.3 & 8.7 & 8.9 & 8.9 & 9.1 & 9.7 & 8.9 & 9.1 & 9.4 \\
log$(M_{*})$ [$\textnormal{M}_{\textnormal{\astrosun}}$] & 10.5 & 10.9 & 11.5 & 10.5 & 10.8 & 12.1 & 10.5 & 10.7 & 11.4 & 10.5 & 10.7 & 12.0 \\
SFR [$\textnormal{M}_{\textnormal{\astrosun}}$/yr] & 0.0 & 0.014 & 0.25 & 0.03 & 0.55 & 6.22 & 0.65 & 3.8 & 52.9 & 0.52 & 6.1 & 276.0 \\
$z$ & - & - & - & 0.022 & 0.090 & 0.211 & - & - & - & 0.017 & 0.118 & 0.250 \\
log$(M_{\textnormal{cold}})$ [$\textnormal{M}_{\textnormal{\astrosun}}$] & 7.7 & 9.1 & 9.9 & - & - & - & 8.4 & 10.1 & 11.2 & - & - & - \\
$^{a}\:M_{\textsc{Hi}}/M_{*}$ & 0.001 & 0.017 & 0.15 & 0.001 & 0.03 & 0.25 & 0.003 & 0.26 & 3.2 & 0.002 & 0.07 & 1.1 \\
$^{b}\:Z_{\textnormal{g}}-Z_{*}$ & -0.84 & 0.07 & 0.30 & -0.78 & -0.19 & 0.87 & -0.12 & 0.17 & 0.54 & -0.37 & 0.26 & 1.9 \\
$M_{\textnormal{bulge}}/M_{*}$ & 0.025 & 1.0 & 1.0 & - & - & - & 0.0 & 0.22 & 1.0 & - & - & - \\
$R_{90,r}/R_{50,r}$ & - & - & - & 1.06 & 2.89 & 3.75 & - & - & - & 1.0 & 2.38 & 6.25\\
Age$_{\textnormal{mw}}$ [Gyr] & 6.9 & 10.7 & 12.3 & - & - & - & 3.6 & 7.2 & 12.0 & - & - & - \\
D$_{n}$4000 & - & - & - & 0.0 & 1.64 & 2.10 & - & - & - & 0.0 & 1.32 & 1.89 \\
log$(M_{\textnormal{BH}})$ [$\textnormal{M}_{\textnormal{\astrosun}}$] & 6.1 & 8.2 & 9.1 & 1.0 & 7.45 & 11.89 & 5.7 & 6.9 & 8.9 & 1.02 & 6.77 & 11.89 \\
$\dot{M}_{\textnormal{cool,net}}$ [$\textnormal{M}_{\textnormal{\astrosun}}$/yr] & 0.0 & 0.0 & 1.7 & - & - & - & 0.0 & 15.0 & 73.0 & - & - & - \\
\hline \hline
\end{tabular}
\caption{The minimum, median and maximum values of the properties analysed for diluting and enriching galaxies from our model sample and Main/$Z_{*}$ observational samples.
\textit{} \newline
$^{a}$ Using the \textsc{Hi}-to-stellar mass fractions obtained via the Z09 scaling relation (Eqn. \ref{eqn:Z09}) for the observational sample, and using $M_{\textnormal{cold}}/M_{*}$ for the model sample.
\textit{} \newline
$^{b}$ For the observational data, only the 61 diluting galaxies and 7,509 enriching galaxies present in the $Z_{*}$ sample are considered. $Z_{*}$ is converted into units of $12+\textnormal{log(O/H)}$ using solar metallicity and oxygen abundance values from \citet{A09} accordingly: $Z_{*}-\textnormal{log}(0.0134)+8.69$.
}
\label{tab:Model_Obs_Comp}
\end{table*}

\subsection{Two classes of massive galaxy in the SDSS} \label{sec:Two classes of real massive galaxy}
When focusing on massive galaxies ($M_{*} \geq 10^{10.5} \textnormal{M}_{\textnormal{\astrosun}}$), we have again selected two sub-samples; the first contains high-SFR, high-$Z_{\textnormal{g}}$ galaxies ($\textnormal{log sSFR} \geq -10.8\: \textnormal{yr}^{-1}$ and $Z_{\textnormal{cold}} \geq 8.85$), and the second contains low-SFR, low-$Z_{\textnormal{g}}$ galaxies ($\textnormal{log sSFR} \leq -10.7\: \textnormal{yr}^{-1}$ and $Z_{\textnormal{cold}} \leq 8.9$). These two regions are marked-out by the red and black dashed lines in the bottom panel of Fig. \ref{fig:SDSS_NumDensity}, respectively. To mimic the terminology used for the model sample, we also refer to these as \textit{enriching} and \textit{diluting} galaxies. However, we emphasise that it is \textit{not} a foregone conclusion that these galaxies are the direct analogues of those in our model, and that it is the purpose of this paper to determine whether this could be the case.

There are 28,838 galaxies in the enriching sub-sample, and 269 galaxies in the diluting sub-sample. Due to the different parameter space coverage, the exact values of sSFR and $Z_{\textnormal{g}}$ used for selection are different from the model sample. However, in both cases, we have attempted to select massive galaxies with `typical' or enhanced star formation for our enriching sub-sample ($\textnormal{i.e.}$ systems on or above the main sequence of star-forming galaxies, e.g. \citealt{E11}), and the low-SFR, low-metallicity tail of the distribution for our diluting sub-sample. In the case of the observational sample, this low-SFR, low-$Z_{\textnormal{g}}$ tail is less extended due to removal of galaxies with low SNR or which host AGN. Therefore, we have chosen a higher upper limit on sSFR for observed diluting galaxies than in the model sample, in order to recover a statistically significant number of galaxies. As with the model, small changes to the regions chosen do not affect our results.

\begin{figure*}
\centering
\includegraphics[totalheight=0.35\textheight, width=0.58\textwidth]{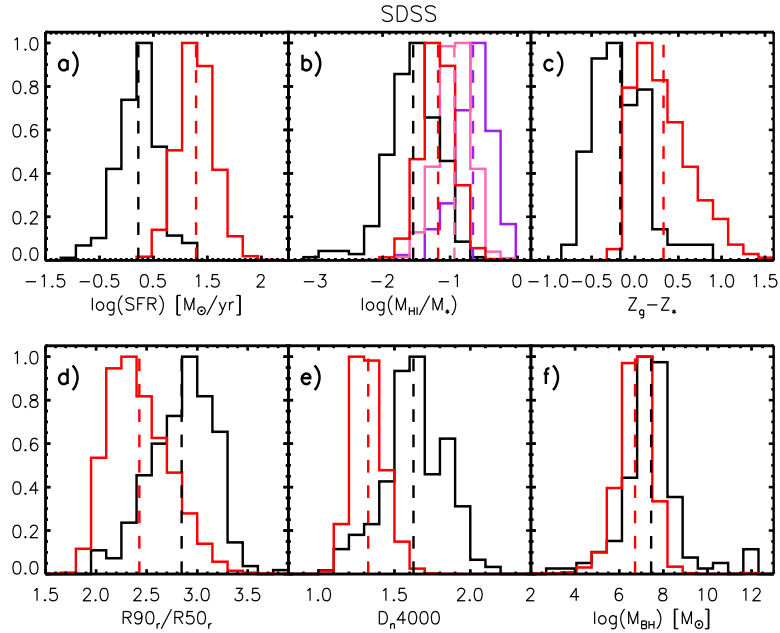}
\caption{The distribution of SFR (panel A), gas-to-stellar mass ratio (from the Z09 scaling relation, panel B), metallicity difference (panel C), concentration index (panel D), D$_{n}$4000 (panel E), and central black hole mass (panel F) for diluting galaxies (black) and enriching galaxies (red) from the observational samples. The pink and purple distributions in panel B show the H\textsc{i}-to-stellar mass ratios for the NUV-r sample and H\textsc{i}-detected sample, respectively. Median values are given by the dashed lines for each distribution.}
\label{fig:SDSS_High-LowZ_hists}
\end{figure*}

Fig. \ref{fig:SDSS_High-LowZ_Mdists} shows the stellar mass distribution for the observational enriching (red) and diluting (black) sub-samples. There is little difference between the distributions for these two sub-samples, meaning that there is no intrinsic mass dependence affecting the results. We have also checked that the diluting galaxies do not exhibit excess star formation in their central regions relative to enriching galaxies from light reprocessed by dust, by comparing their magnitudes around 12 and 22 $\mu$m from the Wide-field Infrared Survey Explorer (WISE).

The key finding of this work is that \textit{all} the signatures of post-merger dilution seen in the semi-analytic model at $z=0$ are also found in our SDSS sample. Fig. \ref{fig:SDSS_High-LowZ_hists} shows that real, `diluting' galaxies have lower SFR (panel A), $M_{\textsc{Hi}}/M_{*}$ (panel B), and $Z_{\textnormal{g}}-Z_{*}$ (panel C), than enriching galaxies, as well as larger $R_{90,r}/R_{50,r}$ (panel D), older ages (panel E), and larger $M_{\textnormal{BH}}$ (panel F). A comparison of the statistical properties of diluting and enriching galaxies between the model and observations is also provided in Table \ref{tab:Model_Obs_Comp}.

It should be noted here that considering the \textit{absolute} values of $Z_{\textnormal{g}}-Z_{*}$ in the semi-analytic model and observations should be treated with caution, as they are sensitive to the set of solar abundances assumed. However, the fact that diluting galaxies typically have lower $Z_{\textnormal{g}}-Z_{*}$ \textit{relative} to enriching galaxies in both the model and observations is a significant result. Also, the fact that both enriching and diluting galaxies in the SDSS have similar median $M_{*}$ and $Z_{*}$ (8.83 and 8.89 in units of $12+\textnormal{log(O/H)}$, respectively) supports the dilution scenario interpretation.

\begin{figure}
\centering
\includegraphics[totalheight=0.25\textheight, width=0.38\textwidth]{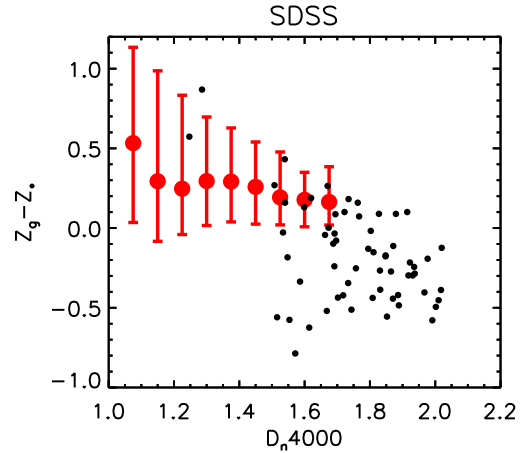}
\caption{The relation between D$_{n}$4000 (a proxy for time since the last starburst) and $Z_{\textnormal{g}}-Z_{*}$ for our observational $Z_{*}$ sample. Red points represent the median $Z_{\textnormal{g}}-Z_{*}$ in bins of D$_{n}$4000 for the `enriching' galaxies. Error bars indicate the 16th and 84th percentiles. Black points represent individual `diluting' galaxies. The oldest diluting galaxies have the lower values of $Z_{\textnormal{g}}-Z_{*}$ than the youngest diluting galaxies, indicating dilution of the gas phase \textit{after} a bout of star formation.}
\label{fig:SDSS_age-Zdiff}
\end{figure}

The red, pink and purple histograms in panel B of Fig. \ref{fig:SDSS_High-LowZ_hists} represent the $M_{\textnormal{H\textsc{i}}}/M_{*}$ distribution for massive, enriching galaxies from the main sample (using the Z09 scaling relation), the NUV-r sample (using the \citealt{C12b} scaling relation) and H\textsc{i}-detected sample (using ALFALFA and GASS measurements), respectively. It is encouraging that all three methods for estimating $M_{\textnormal{H\textsc{i}}}$ show that enriching galaxies have higher gas fractions than diluting galaxies.

We emphasise here that, although it is not surprising to see low-SFR, massive galaxies with low gas fractions, high concentrations, and older ages, it \textit{is} surprising that such galaxies also have low $Z_{\textnormal{g}}$ and low $(Z_{\textnormal{g}}-Z_{*})$. This suggests that these galaxies could be undergoing dilution similar to that seen in some massive galaxies in our semi-analytic model.

Changes to the selection criteria for the whole Main sample do not affect the conclusions drawn for these two classes of massive galaxy. More stringent cuts simply decrease the sample sizes. For example, increasing the minimum SNR for the H$\alpha$, H$\beta$ and [N\textsc{ii}] lines to 10 removes low-sSFR galaxies, and therefore decreases the size of the diluting sub-sample by $\sim 60$ per cent. Lowering the maximum redshift to 0.1 removes some high-SFR galaxies, therefore reducing the enriching sub-sample size by $\sim 66$ per cent. Increasing the minimum fibre-to-total light ratio also reduces the diluting sub-sample significantly. However, the dichotomy seen between the two sub-samples remains strong despite such changes.

\begin{figure*}
\centering
\begin{tabular}{c c}
\vspace{-8.0 mm}
\includegraphics[totalheight=0.25\textheight]{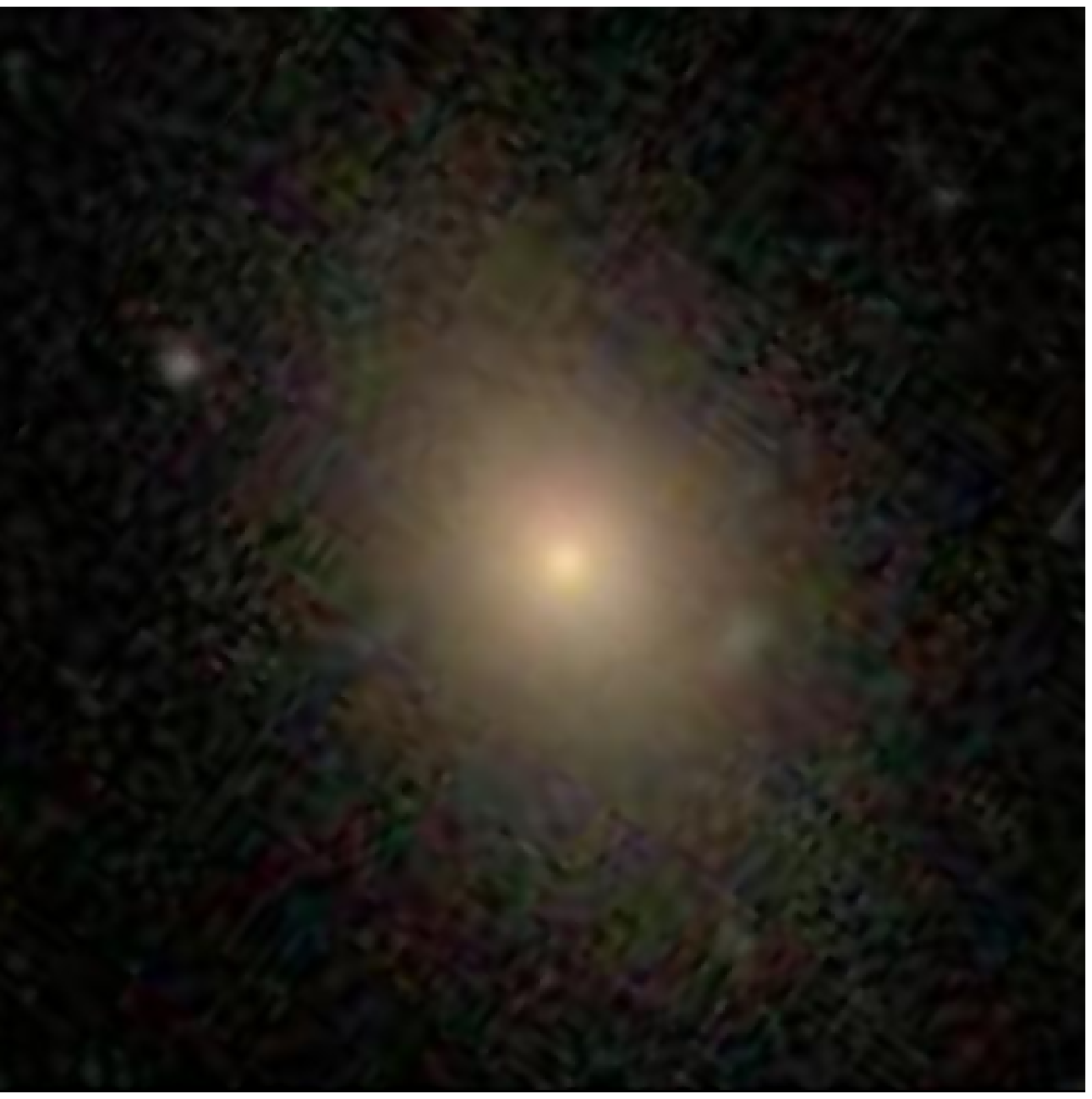} &
\includegraphics[totalheight=0.3\textheight, angle=-90., origin=c]{Fig17b} \\
\vspace{-8.0 mm}
\includegraphics[totalheight=0.25\textheight]{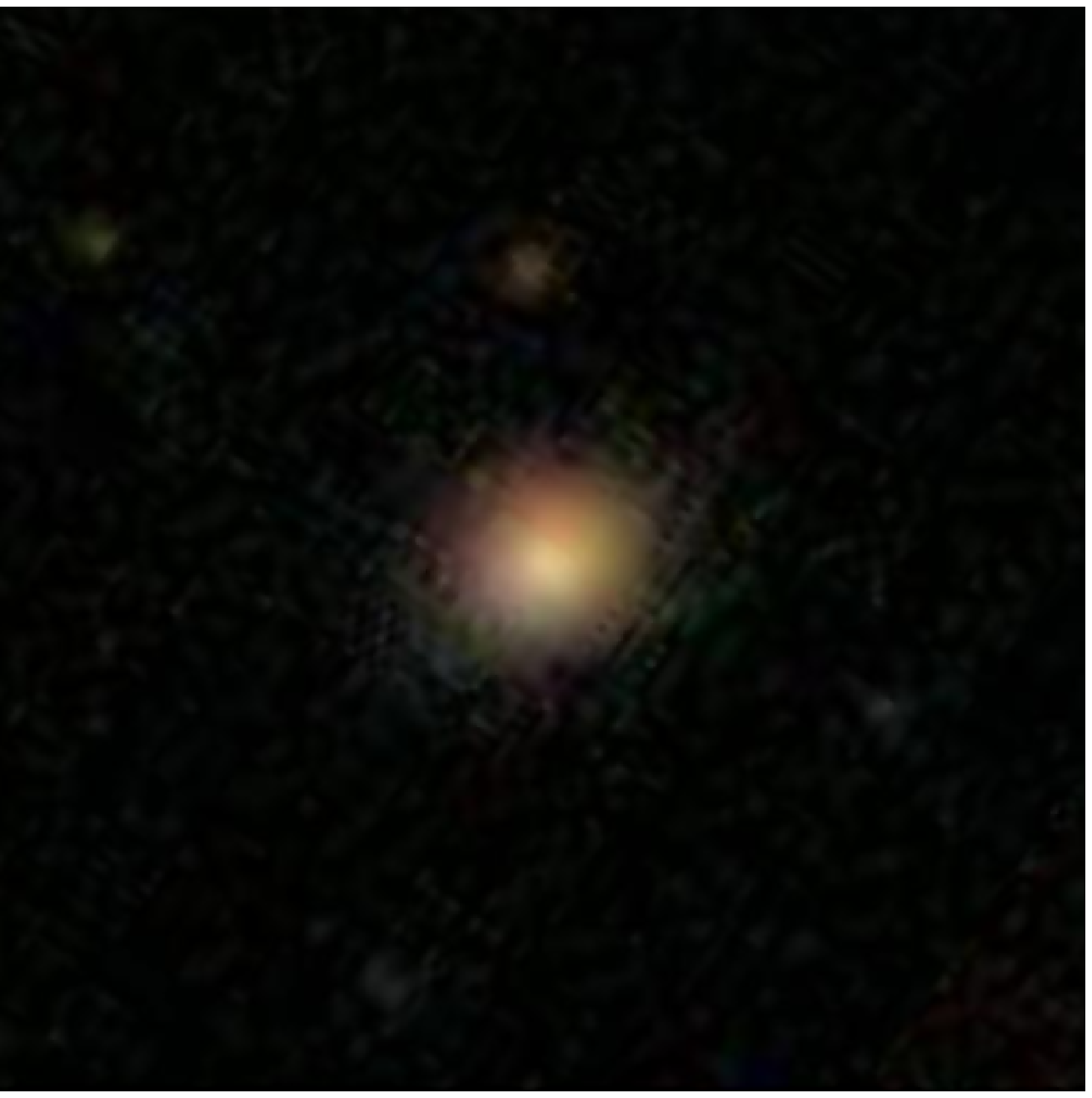} &
\includegraphics[totalheight=0.3\textheight, angle=-90., origin=c]{Fig17d} \\
\vspace{-8.0 mm}
\includegraphics[totalheight=0.25\textheight]{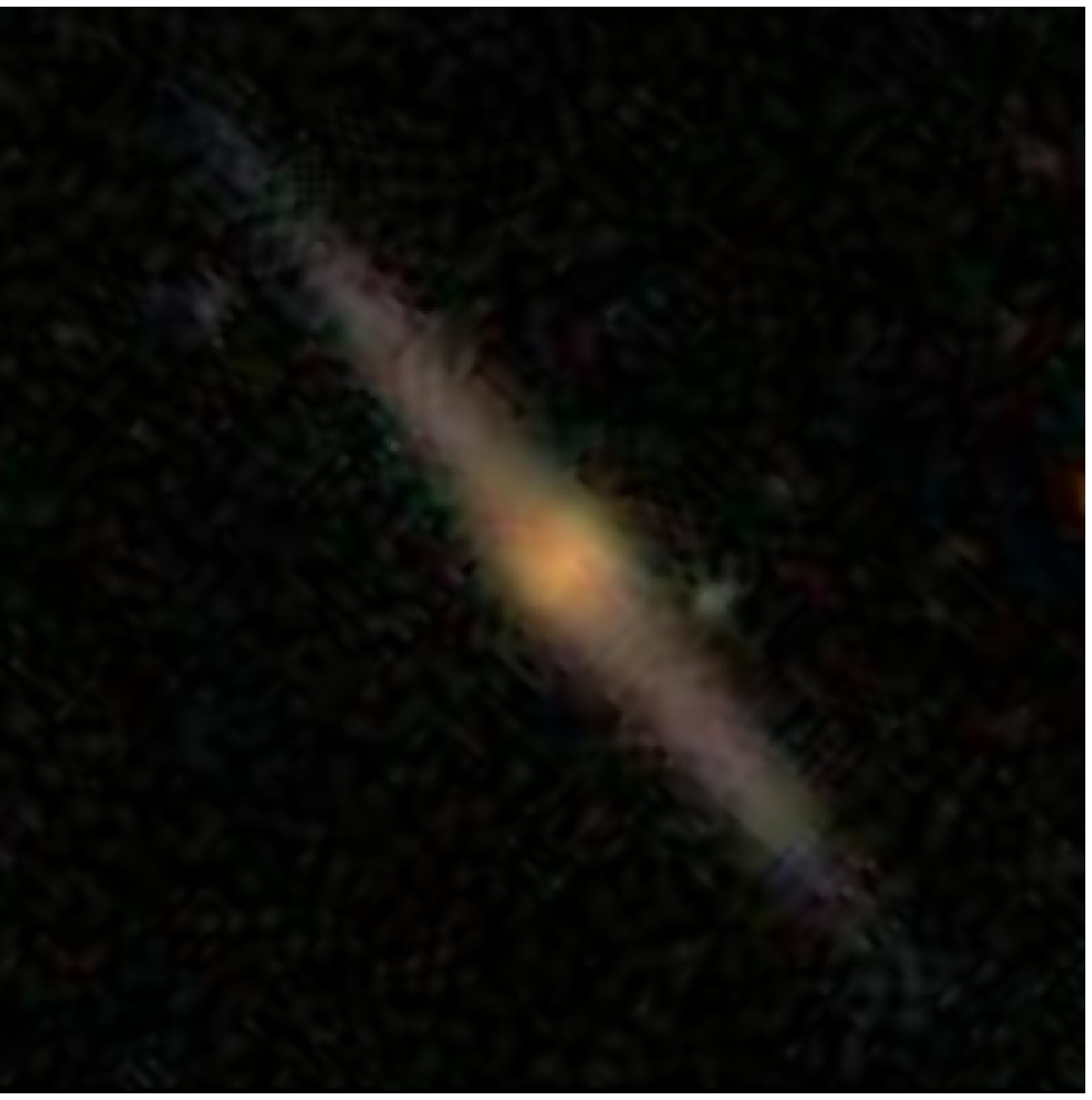} &
\includegraphics[totalheight=0.3\textheight, angle=-90., origin=c]{Fig17f} \\
\vspace{-8.0 mm}
\includegraphics[totalheight=0.25\textheight]{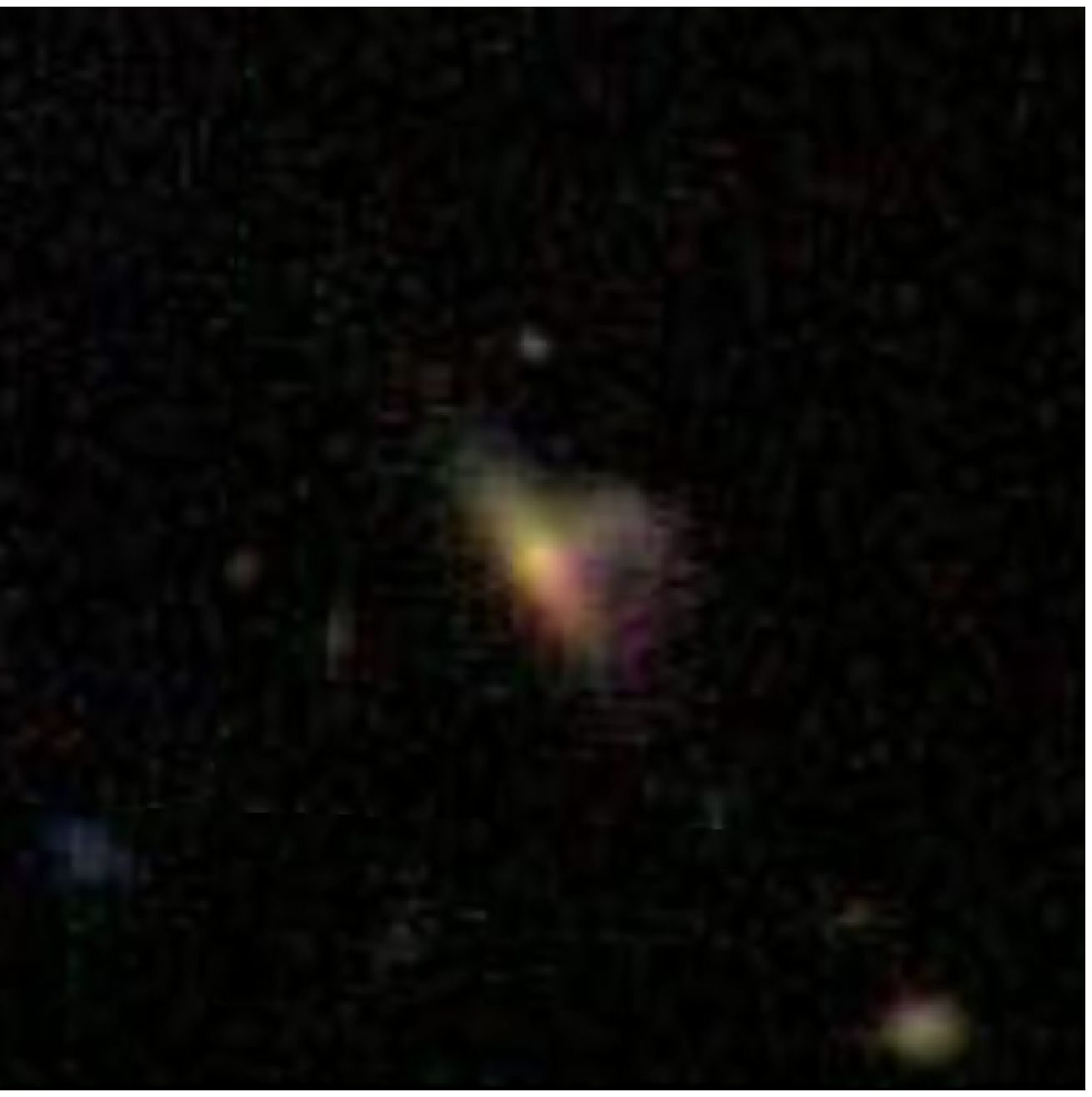} &
\includegraphics[totalheight=0.3\textheight, angle=-90., origin=c]{Fig17h} \\
\end{tabular}
\caption{Four example objects from our SDSS-DR7 sub-sample of diluting galaxies. Thumbnail images and optical spectra are shown for two elliptical galaxies, an edge-on disc and an interacting system.}
\label{fig:SDSS_image_collage_wSpectra}
\end{figure*}

Importantly, all the trends described above also hold when using the strong-line-ratio-based metallicities of \citet{M10}, even though this method predicts higher metallicities for galaxies below $Z_{\textnormal{g}}\sim 9.1$ than the Bayesian method (see YKG12, \S 4.1). For example, the mean value of the \textsc{Hi}-to-stellar mass ratio for diluting galaxies when selecting by strong-line-ratio-based metallicity is $\textnormal{log}(M_{\textnormal{H\textsc{i}}}/M_{*})_{\textnormal{Z09}} = -1.6$. This is still 0.42 dex lower than the mean value for enriching galaxies using the same selection criteria. Also, the difference in mean $(Z_{\textnormal{g}}-Z_{*})$ between diluting and enriching galaxies when using the strong-line diagnostic to obtain $Z_{\textnormal{g}}$ is still 0.5 dex. We therefore consider the \textit{relative} properties of enriching and diluting galaxies to also be robust to the metallicity diagnostic chosen.

Fig. \ref{fig:SDSS_age-Zdiff} shows the relation between D$_{n}$4000 and $Z_{\textnormal{g}}-Z_{*}$ for massive galaxies in our observational $Z_{*}$ sample. This can be compared to Fig. \ref{fig:LGals_age-Zdiff} for the model. Although the number of observed diluting galaxies (black points) with reliable stellar metallicities is relatively small (61 galaxies), there is still a clear trend present -- the oldest diluting galaxies tend to have the lowest $Z_{\textnormal{g}}-Z_{*}$. If anything, the anti-correlation between age and $Z_{\textnormal{g}}-Z_{*}$ is even stronger in the observations than in the model, which is to be expected when using D$_{n}$4000 as a proxy for age, as it measures the time since the last bout of star formation (see $\textnormal{e.g.}$ \citealt{K03a}). This is further indirect evidence of a gradual dilution of the ISM after a starburst has occurred.

It is also interesting to note that many of the diluting galaxies in our observational sample have early-type morphologies. From visual inspection of the SDSS-DR7 optical thumbnail images alone, $\sim50$ per cent appear to be elliptical in shape and lack significant blue emission. This is supported by the higher average concentration index in the diluting sub-sample (Fig. \ref{fig:SDSS_High-LowZ_hists}, panel D), and reflects the large bulge-to-total stellar mass ratios seen for diluting galaxies in our model (Fig. \ref{fig:LGals_SSFR-Z_maps}, panel E). A further $\sim 32$ per cent of the observational sub-sample appears to be edge-on disc galaxies. These are likely assigned low SFRs and low gas-phase metallicities due to their greater optical thickness, which reduces the amount of emission observed from their galactic centres. $\sim 16$ per cent of the diluting galaxies are either currently interacting or of uncertain morphology. Only the final $\sim 2$ per cent is made up of objects that appear to be nearly face-on disc galaxies with some blue emission. We note, of course, that by-eye classification using only low-resolution optical images can only give a rough indication of the typical morphologies for a sample of nearby galaxies.

Fig. \ref{fig:SDSS_image_collage_wSpectra} shows optical images of four representative galaxies in the diluting sub-sample, along with their SDSS spectra.\footnote{Images and spectra obtained from \textit{http://cas.sdss.org/astrodr7/en/tools/chart/list.asp}} Although most emission lines are not particularly strong, [N\textsc{ii}] and H$\alpha$ are well detected in all four spectra and could be dominating the metallicity estimates. To test the significance of the [N\textsc{ii}]$\lambda$6584 line, our analysis was re-run using Bayesian metallicity estimates that do not require [N\textsc{ii}] (or [S\textsc{ii}]) in the fit. Although this does lower the $Z_{\textnormal{g}}$ estimate for some massive galaxies, it doesn't affect the relative median $Z_{\textnormal{g}}-Z_{*}$ values between the diluting and enriching sub-samples. YKG12 have also shown that removing [N\textsc{ii}] from the estimation doesn't affect the positive correlation between SFR and $Z_{\textnormal{g}}$ at high $M_{*}$. We also note that the majority of galaxies exhibiting low ionisation nuclear emission line regions (LINERs) have already been removed from our sample via the AGN cut described in \S \ref{sec:Main sample}. The [N\textsc{ii}]/H$\alpha$ ratio would not provide an accurate estimate of the host galaxy's $Z_{\textnormal{g}}$ in such systems.

When removing all galaxies with disc-like morphologies from the diluting sub-sample, a slight decrease in SFR, $M_{\textnormal{H\textsc{i}}}/M_{*}$, and $Z_{\textnormal{g}}-Z_{*}$, along with a slight increase in $R_{90,r}/R_{50,r}$, D$_{n}$4000, and $M_{\textnormal{BH}}$ is seen, as would be expected. For example, the median SFR drops by $0.09 M_{\textnormal{\astrosun}}/$yr, the median D$_{n}$4000 rises by 0.03, and the median $M_{\textnormal{BH}}$ rises by 0.24 dex. 

\textit{} \newline
To conclude this section, we can say that all the signatures of post-merger dilution seen in our model sample are also found in low-SFR, low-$Z_{\textnormal{g}}$, massive galaxies in the SDSS. This is strong, indirect evidence for claiming that real galaxies have also undergone a gradual dilution of their gas phase, after the truncation of continuous star formation following a merger-induced starburst. However direct measurements of metal-poor gas infall onto these galaxies have not yet been made, and so further observational studies are required to confirm or deny the conclusions drawn from this work.

\section{Comparisons to other works} \label{sec:Comparisons to other works}
\subsection{Accretion onto elliptical galaxies} \label{sec:Accretion onto elliptical galaxies}
There is already a class of spheroidal galaxy identified as possibly undergoing accretion of metal-poor gas, known as polar ring galaxies (PRGs) \citep{SWR83}. PRGs tend to exhibit extended rings or discs of H\textsc{i} gas, dust and sometimes young stars, lying perpendicularly to the equatorial plane of the central spheroid, with the kinematics of the two components decoupled \citep{W90}. One explanation for the formation of the rings is the later accretion of cold gas, which is either stripped from nearby or merged satellites ($\textnormal{e.g.}$ \citealt{RS97,BoCo03,H09}), or accreted from cosmic filaments ($\textnormal{e.g.}$ \citealt{M06,S10,SI13}). However, a scenario where a major, gas-rich merger forms both the central spheroid and the outer polar ring together is also possible ($\textnormal{e.g.}$ \citealt{B98,I02a}).

\citet{M11} have recently compiled the Sloan-based Polar Ring Catalogue (SPRC) of 275 nearby PRGs also observed by the SDSS. We find that 61 of these PRGs are also present in our Main observational sample, 20 of which have $\textnormal{log}(M_{*}) \geq 10.5$. Of these 20 massive PRGs, only 2 are present in our diluting sub-sample: SPRC-37 and SPRC-183. The first has $\textnormal{log(sSFR)}=-11.17\: \textnormal{yr}^{-1}$ and $Z_{\textnormal{g}}=8.5$, and the second has $\textnormal{log(sSFR)}=-11.57\: \textnormal{yr}^{-1}$ and $Z_{\textnormal{g}}=8.5$.

Interestingly, these two systems have the first and third oldest ages of all the PRGs in our Main sample, with D$_{n}4000=1.79$ and $1.67$, respectively (the median value for the diluting sub-sample is 1.64). In addition, they also have the first and third lowest metallicity differences, with $Z_{\textnormal{g}}-Z_{*}=-0.13$ and $-0.52$, respectively (the median value for the diluting sub-sample is -0.19). This could be an indication that metal-poor gas is \textit{gradually} accreted onto such systems over time, again supporting the dilution scenario described in this work.\footnote{We note here that metal-poor gas in `face-on' polar rings may not be considered in the estimation of $Z_{\textnormal{g}}$, due to the narrow 3-arcsecond fibres used by the SDSS. Therefore, these objects may have lower overall $Z_{\textnormal{g}}$ than is observed.}

The fact that only two of our observed diluting galaxies are confirmed as PRGs in the SPRC suggests that this is not the main mechanism by which elliptical galaxies accrete metal-poor gas. Indeed, the presence of an AGN, which is often the case for our model diluting sub-sample, would likely preclude the accretion and cooling of gas from filaments (although not necessarily from satellite stripping or mergers).

\citet{R13} have recently suggested that the BCG of Abell 1664, which hosts an AGN, could be undergoing inflow of two molecular gas clumps, which could settle into a disc over several hundreds of megayears (although, they also point out that this material could be part of an outflow, driven by AGN feedback). Such a mode of accretion is much more common for diluting galaxies in our semi-analytic model than smooth infall and cooling from the intergalactic medium (see Fig. \ref{fig:GasFracPaper_LGals_CoolvsAcc}). The advent of the Atacama Large Millimeter Array (ALMA) survey should hopefully facilitate many more observations of infalling molecular gas onto massive ellipticals in the future.

Similarly, \citet{Hu11} have found a massive ($M_{\textnormal{bulge}} \lesssim 3.4\times 10^{10} \textnormal{M}_{\textnormal{\astrosun}}$) QSO-host galaxy at $z\sim 0.2$ with a large black hole ($M_{\textnormal{BH}}\sim 3\times 10^{8} \textnormal{M}_{\textnormal{\astrosun}}$) and very low gas-phase metallicity ($Z_{\textnormal{g}} \lesssim 8.4$). The low metallicity in this system is believed to be due to dilution, either from accretion of metal-poor gas stripped from satellites or smooth accretion from the ambient gas reservoir. \citet{H12} have further found that bulge-dominated, QSO-host galaxies typically have lower $Z_{\textnormal{g}}$ than their disc-dominated counterparts, and that both have lower $Z_{\textnormal{g}}$ than `non-active' star-forming galaxies of the same mass (although \citealt{SL13} suggest that $Z_{\textnormal{g}}$ can be under-estimated in active galaxies). These low-$Z_{\textnormal{g}}$ objects are quite distinct from the majority of massive AGN hosts which have super-solar $Z_{\textnormal{g}}$ ($\textnormal{e.g.}$ \citealt{HF93,GHK06}), but they do exhibit properties seen in diluting galaxies in our semi-analytic model.

\subsection{Interacting galaxies} \label{sec:Interacting galaxies}
Around $\sim 8$ per cent of the massive, `diluting' galaxies in our observational sample appear, from their SDSS images, to be interacting. \citet{K10} have shown that four close pairs of galaxies in the local Universe have lower-than-expected central $Z_{\textnormal{g}}$, likely due to rapid migration of metal-poor gas from larger radii into the centres of each galaxy during the interaction (see also \citealt{KGB06,R08}). Such a process is most effective when the interacting galaxies are of similar mass \citep{W06,E08b,MD08}. \citet{Mo10} and \citet{T12} have also shown that such a process occurs in their SPH simulations of equal-mass, interacting, disc galaxies.

One of the close pairs investigated by \citet{K10} is also present in our main observational sample, comprising NGC 3994 and NGC 3995. We find that, although these two galaxies have relatively low central $Z_{\textnormal{g}}$ (9.01 and 8.67, respectively), they also have high log(sSFR) ($-8.28\:\textnormal{yr}^{-1}$ and $-8.41\:\textnormal{yr}^{-1}$), high $M_{\textnormal{H\textsc{i}}}/M_{*}$ (2.55 and 1.40), and young ages (D$_{n}$4000 = 1.16 and 1.01). High SFRs were also found for the majority of the 42 low-$Z_{\textnormal{g}}$, interacting SDSS galaxies studied by \citet{PPS09}. These properties are to be expected for galaxies with metal-poor gas flooding into the central regions inducing a nuclear starburst. However, they are not found in any of the small number of interacting systems in our diluting sub-sample. Such a sudden dilution of the ISM is therefore unlikely to be the sole cause of low $Z_{\textnormal{g}}$ in massive galaxies with disturbed morphologies.

It could be that the interacting galaxies studied by \citet{K10} are in an initial phase of the evolution seen in our model, and will undergo a gradual dilution in the future. However, the simulations of \citet{Mo10} and \citet{T12}, as well as \textsc{L-Galaxies}, show that $Z_{\textnormal{cold}}$ increases again shortly after a merger event, due to metal enrichment from the first supernov\ae{} following the starburst. Therefore, any dilution after this time would not be due to the initial interaction, and would have to occur through a second process (such as accretion), in the absence of strong star formation.

\subsection{The relation between $M_{*}$, $Z_{g}$, and $M_{\textit{HI}}$} \label{sec:The relation between M*, Zg and MHI}
As mentioned in \S \ref{sec:HI scaling relations}, Z09 also studied the dependence of the $M_{*}$-$Z_{\textnormal{g}}$ relation on \textsc{Hi} gas mass fraction. They found that, at fixed $M_{*}$, metal-rich galaxies have lower gas fractions than metal-poor galaxies. This may seem to contradict the findings in this work at high mass. However, these two results are compatible with each other due to the difference in the sample selection. Z09 selected galaxies with log(sSFR) $> -11.0$ yr$^{-1}$, meaning that only 45 per cent of the diluting galaxies we consider in our observational analysis, and none of the diluting galaxies in our model analysis, meet the same criterion.

When we also select a sample of massive galaxies with log(sSFR) $> -11.0$ yr$^{-1}$, we recover similar trends to those found by Z09, for both our observational and model samples. It is the particular class of low-sSFR, diluting galaxies -- that are \textit{not} present in the Z09 sample -- which deviate from these trends by having low $Z_{\textnormal{g}}$ \textit{and} low gas fractions at $z=0$.

The case is similar when comparing our results to those of \citet{Hu13}, who also observed an anti-correlation between gas fraction and gas-phase metallicity in 260 local, late-type galaxies (although this is difficult to determine at high mass, due to the low number of galaxies above $\textnormal{log}(M_{*}) > 10.5 \textnormal{M}_{\textnormal{\astrosun}}$ in their sample). Their selection of galaxies with $\textnormal{log(sSFR)} \geq -10.9$ yr$^{-1}$ also excludes most of the diluting galaxies analysed in this work.

Pleasingly, our findings support the cartoon model prediction described by \citet{LL13}. In that work, it was postulated that massive, low-sSFR, low-$Z_{\textnormal{g}}$ galaxies should have low gas fractions. Our results, using a wide range of observational data and our galaxy formation model, show that this is indeed the case. We further argue that the cause of this trend is a gradual dilution of the gas phase after a merger-induced starburst.

\subsection{The FMR at z=0} \label{sec:The FMR at z=0}
YKG12 have pointed-out that the shape of the FMR at $z=0$ is strongly dependent on the $Z_{\textnormal{g}}$ diagnostic chosen. Consequently, there are conflicting conclusions in the literature as to the true nature of this $M_{*}$-SFR-$Z_{\textnormal{g}}$ relation at high mass \citep{M10,LL10,YKG12,AM12,LL13,Z13,B13}. It is therefore important to find viable physical processes that can \textit{explain} a given correlation between $M_{*}$, SFR and $Z_{\textnormal{g}}$. There now appears to be three observationally-motivated explanations for why some massive galaxies can fall-off the $M_{*}$-$Z_{\textnormal{g}}$ relation to lower metallicities: a) rapid dilution of the central regions of high-SFR, interacting galaxies ($\textnormal{e.g.}$ \citealt{KGB06}), b) the removal of dust and metals from low-SFR, secularly-evolving galaxies by radiation pressure (known as the slow-flow, dust-efflux model, \citealt{Z13b}), and c) gradual dilution of low-SFR, elliptical galaxies in the absence of secular star formation (as proposed in this work). The second of these provides a direct explanation for the relation between $M_{*}$ and dust extinction in low-$z$ galaxies \citep{Z13}. However, it is not certain that the radiation-driven winds would be of higher \textit{metallicity} than that measured in the galaxies' H\textsc{ii} regions. If any (or a combination) of these three mechanisms is happening to a significant number of galaxies at low redshift, then the shape of the FMR at high mass would be more complicated than first assumed. Therefore, the full population of galaxies in the local Universe would not be adequately fit by such a straightforward relation, and caution should be taken when using the FMR to describe galaxies with $M_{*} \gtrsim 10^{10.5} \textnormal{M}_{\textnormal{\astrosun}}$ at low redshift.

\section{Conclusions} \label{sec:Conclusions}
We have compared various physical properties of two classes of massive galaxy in the SDSS-DR7 and Munich semi-analytic model of galaxy formation, \textsc{L-Galaxies}. These two classes are selected by their specific star formation rates and gas-phase metallicities; high-sSFR, high-$Z_{\textnormal{g}}$ systems are labelled as `enriching' galaxies, and low-sSFR, low-$Z_{\textnormal{g}}$ system are labelled as `diluting' galaxies. The following results were obtained from this comparison:

\begin{itemize}
\item Diluting galaxies in the semi-analytic model have higher bulge-to-total mass ratios, mass-weighted ages, and central black hole masses than model enriching galaxies at $z=0$, and lower cold gas masses, gas-to-stellar mass ratios, and differences between their gas-phase and stellar metallicities.

\item These properties are all signatures of the specific evolution undergone by such galaxies -- a gas-rich merger and subsequent starburst, followed by a cessation in secular star formation. A \textit{gradual} dilution of the gas phase then takes place for up to several gigayears, via the accretion of metal-poor, cold gas in clumps and low-mass merging satellites. This gradual dilution drives the positive correlation between SFR and $Z_{\textnormal{cold}}$ seen in massive galaxies at $z=0$ in the model.

\item \textit{All} the signatures of the evolution described above are also seen in low-sSFR, low-$Z_{\textnormal{g}}$, massive galaxies in the SDSS-DR7. Of particular note are their elliptical morphologies, low gas fractions and low $Z_{\textnormal{g}}-Z_{*}$, which suggest dilution of the gas-phase in the absence of star formation. This is strong, indirect evidence that gradual dilution \textit{after} a gas-rich merger event is taking place in some elliptical galaxies in the local Universe.

\item These results suggest an alternative mechanism by which galaxies can fall off the $M_{*}$-$Z_{\textnormal{g}}$ relation to lower metallicities. In this scenario, galaxies remain at low $Z_{\textnormal{g}}$ for a longer period than possible via rapid dilution (and re-enrichment) \textit{during} mergers and interactions.

\item The \textit{positive} correlation found between SFR and $Z_{\textnormal{g}}$ in the local Universe shows that current formulations of the FMR (which assume a weak \textit{anti}-correlation between SFR and $Z_{\textnormal{g}}$ at high mass) do not accurately represent the whole galaxy population.
\end{itemize}

We close by highlighting some important limitations of the observational analysis in this work. First, the gas-phase metallicities used are measured from light falling within the 3-arcsecond aperture of the SDSS fibres. This covers the inner $\sim 1$ to 9 kpc of galaxies in our Main sample. Gas lying at larger radii will therefore not be included in the metallicity estimates.

Second, a significant amount of high-$M_{*}$, low-$Z_{\textnormal{g}}$ galaxies are missing from our observational analysis because of poorly-constrained estimates of their key physical properties. Future observations of such galaxies' gas content from near-IR absorption lines in the CGM ($\textnormal{e.g.}$ \citealt{P13}), and gas-phase metallicity from temperature-insensitive emission lines in the far-IR ($\textnormal{e.g.}$ \citealt{C13}) would greatly help us probe this important part of the galaxy population. Likewise, the wealth of information available for (potentially) low-$Z_{\textnormal{g}}$ AGN-host galaxies remains untapped, due to the contamination of their spectra by emission from the nucleus. Continued measurements of $Z_{\textnormal{g}}$ from their narrow-line regions ($\textnormal{e.g.}$ \citealt{H12}), or of the molecular gas in and around these objects ($\textnormal{e.g.}$ \citealt{R13}), would open-up these systems for future analysis.

\section*{Acknowledgments} \label{sec:Acknowledgements}
The authors would like to thank Barbara Catinella, Silvia Fabello, Anna Gallazzi, Bruno Henriques, Bernd Husemann, Chervin Laporte, Maritza Lara-L\'{o}pez, Mei-Ling Huang, Jing Wang, and Jabran Zahid for their help, advice and inspiration during the undertaking of this work. R. M. Y. also acknowledges the financial support of the Deutsche Forschungsgesellshaft (DFG).

\section*{Appendix A: Cross-matching SDSS and GASS samples} \label{sec:Appendix A}
In the GASS data, objects are linked to their SDSS counterparts via an SDSS identifier, which is simply the concatenation of the right ascension (\textit{ra}) and declination (\textit{dec}) of the object in hexadecimal format, such that, SDSS ID = Jhhmmss.ss+ddxxyy.y. In these IDs, J indicates the use of the J2000 standard equinox, \textit{ra} = hhmmss.ss in hours, minutes and seconds, and \textit{dec} = ddxxyy.y in days, arcminutes (xx) and arcseconds (yy.y). The + sign before the declination indicates that the object lies in the northern hemisphere (there are no southern hemisphere objects in GASS). Once decomposed from the SDSS ID, \textit{ra} and \textit{dec} from GASS can be compared to those from any SDSS data release following the straightforward conversion to degrees; \textit{ra}$_{\textnormal{deg}} = (360/24) \cdot [\textnormal{hh} + (\textnormal{mm}/60) + (\textnormal{ss.ss}/3600)]$ and \textit{dec}$_{\textnormal{deg}} = \textnormal{dd} + (\textnormal{xx}/60) + (\textnormal{yy.y}/3600)$.

\end{document}